\documentclass[aps,prb,showpacs,twocolumn,longbibliography, superscriptaddress]{revtex4-1}

\usepackage[dvips]{graphicx}
\usepackage{amsmath,amssymb}
\usepackage{amsfonts}
\usepackage[table,usenames,dvipsnames]{xcolor}
\usepackage[retainorgcmds]{IEEEtrantools}

\usepackage{placeins}  
\usepackage[normalem]{ulem}      

\begin{document}

\newcommand{\Efl}{E_{\text{fl}}}
\newcommand{\Ib}{I_{\text{b}}}
\newcommand{\Ediss}{E_{\text{diss}}}
\newcommand{\Done}{\texttt{D1}}
\newcommand{\Dtwo}{\texttt{D2}}
\newcommand{\Dthr}{\texttt{D3}}
\newcommand{\DoneJ}{\texttt{D1J}}
\newcommand{\DtwoJ}{\texttt{D2J}}
\newcommand{\DthrJ}{\texttt{D3J}}
\newcommand{\uGHz}{\,\text{GHz}}
\newcommand{\umicroA}{\,\mu\text{A}}
\newcommand{\uOhm}{\,\Omega}
\newcommand{\upH}{\,\text{pH}}
\newcommand{\uzJ}{\,\text{zJ}}
\newcommand{\uaJ}{\,\text{aJ}}
\newcommand{\Eq}[1]{Eq.~\eqref{#1}}
\newcommand{\Eqs}[1]{Eqs.~\eqref{#1}}
\newcommand{\sech}{\;\text{sech}} 
\newcommand{\diff}{\text{d}}

\title{Boosting fluxons for ballistic-logic power using an Aharonov-Casher ring} 

\author{W.~Wustmann}
\affiliation{The Laboratory for Physical Sciences at the University of Maryland, 
College Park, MD 20740, USA}

\author{K.D.~Osborn}
\email{kosborn@umd.edu (corresponding author)}
\affiliation{The Laboratory for Physical Sciences at the University of Maryland, 
College Park, MD 20740, USA}
\affiliation{The Joint Quantum Institute, University of Maryland, 
College Park, MD 20742, USA}
\affiliation{The Quantum Materials Center, University of Maryland,
College Park, MD 20742, USA}

\date{\today}

\begin{abstract}
Superconducting logic is fast and energy-efficient relative to CMOS, 
but also fundamental studies 
are needed to scale up circuits for greater utility.
Recently, ballistic shift registers for single-flux quanta (SFQ) bits were shown in simulations to allow high-efficiency superconducting gates. 
However, these gates are unpowered such that 
the bits slow after each gate operation, and thus
only a short sequence of gates is possible without added power.
Here we theoretically show that a circuit based on an Aharonov-Casher ring can power these shift registers by boosting the bit velocity to a constant value, despite their unusual bit states constituted by two polarities of SFQ. 
Each bit state is forced into a different ring arm and then accelerated as part of the operation.
The circuit dynamics depend on various circuit parameters and choices of how to merge the bit-state paths. 
One design from each merge design choice is proposed to possibly enable scaling up to an array of gates by adding serial biasing in a relatively simple way.
We find adequate performance for ballistic logic in terms of boosted velocity, energy efficiency, and parameter margins. We also discuss the circuit's classical barriers, which relate to the Aharonov-Casher effect in a different parameter regime.  
\end{abstract}


\maketitle 

\section{Introduction}

Superconducting digital logic \cite{Holmes2013} has the potential 
for high-performance computing due to its high energy efficiency and speed relative to CMOS logic, and is partially motivated by the large power consumed by today's computer networks and data centers. 
This type of logic uses single flux quanta (SFQ) as bits in arithmetic logic units \cite{ArithmeticLogic1, ArithmeticLogic2} and processors \cite{Processor1, Processor2, Processor3}. SFQ logic is used in analog-to-digital converters and broadband digital receivers \cite{Receiver1, Receiver2, Receiver3, Receiver4}, and Josephson voltage standards are related through their voltage pulses \cite{VoltageStandard1,  VoltageStandard2, VoltageStandard3, VoltageStandard4, VoltageStandard5}.

SFQ logic sometimes refers to the pioneering logic family named Rapid Single Flux Quantum (RSFQ) logic \cite{LikMukSem1985, PolonskyETAL1993}, 
and its variants which are made for higher efficiency \cite{YoshikawaBias, FujimakiBias, KirETAL2011, VolETAL2013, MukhanovETAL2014}. 
In RSFQ, an SFQ of a fixed polarity represents the bit state ``1'', and its absence the ``0'', and dc-bias currents power data and clock SFQ along specified paths to execute the logic.
In contrast, ac-powered logic families, named Reciprocal Quantum Logic \cite{HerrETAL2011} and Adiabatic Quantum Flux Parametron \cite{YoshikawaETAL2013_10zJ, YoshikawaETAL2021}, are studied for their higher efficiency with different biasing. 
However, the most extreme methods for high efficiency borrow concepts from thermodynamic reversibility, which in principle allow the energy cost of some adiabatic logic \cite{parametricquantron1985, RenSemETAL2009, YoshikawaETAL2013} and ballistic logic \cite{ OsbWus2020_CNOT,OsbWus2023_BSR, Frank2017, FrankETAL2019_ISEC} 
to scale below $\ln(2) k_B T$ per gate.

A recent thrust for SFQ logic aims to scale up the operating gate count in circuits, 
e.g.~ref.~\onlinecite{TolpygoETAL2015, Tzimpragos}.
Related to this,
small bit-storage capacity memory is studied as a register~\cite{Register1, Register2, Register3},
which in general is made from a sequence of 1-bit latches named shift registers~\cite{ShiftRegister}.
Larger memory is studied as RAM which tests SFQ logic in a Von-Neumann architecture~\cite{RAM1, RAM2, RAM3}. 
Traditionally, bias current (in RSFQ) is sourced to the chip ground plane 
yielding a combined current that increases with the number of gates and imposes a limitation analogous to power-density limitations in CMOS electronics~\cite{Mukhanov2011}.
A rarely implemented method to reuse bias current, called serial biasing or bias-current recycling \cite{CurrentRecycling0, CurrentRecycling4, CurrentRecycling1, CurrentRecycling3, CurrentRecycling2b} 
offers one possible solution to this scaling problem.

For our ballistic logic type called Reversible Fluxon Logic (RFL) \cite{WusOsb2020_RFL, LiuqiDesign, OsbWus2020_CNOT}, we have recently simulated ballistic shift registers (BSRs) \cite{OsbWus2023_BSR}, which are gates that allow fully asynchronous inputs as long as a minimum time delay exists between input bits. 
RFL uses degenerate (equal energy) bit states, where each polarity represents a state. 
The logic is considered ballistic because the SFQs move ballistically in long Josephson junctions (LJJs) as transmission lines and because 
some gates are solely powered by input-bit inertia. 
However, to make BSRs practical, a power source is needed to accelerate bits between gates or pairs of gates.

\begin{figure}[b]
\includegraphics[width=\columnwidth]{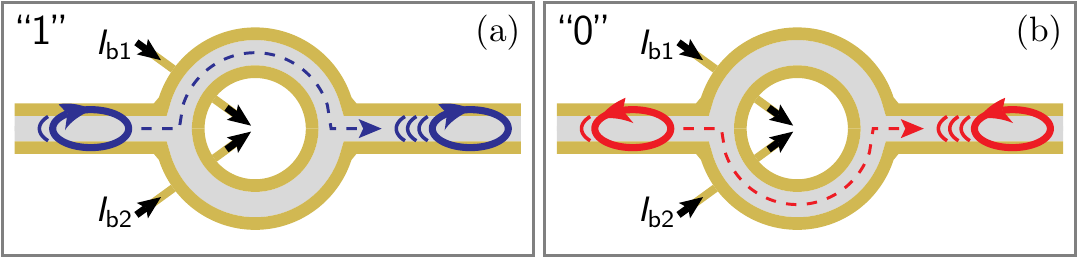}
\caption{
Illustration of boost operation for fluxons of (a) positive (``1'') 
and (b) negative (``0'') polarity (bit state). 
The Aharonov-Casher (A-C) ring is built from two short LJJ arms and is connected to two LJJ ports for input and output.
Bias currents $I_{\text{b}1} = I_{\text{b}2} = \Ib$ 
are applied in both arms near the input branch point, 
and the bias generates a clockwise (anticlockwise) force on an incoming fluxon (antifluxon). 
This force allows the fluxon to enter one of the arms, as determined by the force, and accelerates it. The critical currents of the ring LJJs are increased near the output port to allow the fluxon to exit the A-C ring in forward transmission.
}
\label{fig:boosteroperations} 
\end{figure}

In this theoretical proposal, we re-purpose the ring structure from early Aharonov-Casher (A-C) proposals \cite{ACeffect0, ACeffect1, ACeffect2, ACeffect3}. 
The A-C effect, i.e. the self-interference of a flux-carrying quantum particle 
when taking multiple paths,
occurs in various superconducting systems 
\cite{ManucharyanETAL2012, PopETAL2012, EliETAL1993, BellETAL2016, Topological2, Topological3, YamashitaETAL2003}
in which flux quantization naturally plays a pivotal role.   
In LJJs, SFQ exist in the form of fluxons (flux solitons)
which, however, behave highly classical under typical material characteristics \cite{AveRabSem2006},
and for that reason, no significant self-interference effects are expected 
in our Nb-based LJJ ring. 
Nonetheless, we find that the same A-C ring structure (with input and output ports) is 
useful to power RFL. 
RFL cannot simply use a bias current applied to an LJJ 
to restore bit energy (a minimum value is required in gate operations), 
because it would accelerate one of the two states of a bit but not the other.
However, by applying current biases in the LJJ ring, we can force the two states of the bit along its two different arms, and accelerate both bit states forward.
In simulations, we find that we can boost fluxons to a velocity that is insensitive to the input velocity 
using only two dc-bias currents, modified LJJs, and damping resistors. 
Furthermore, we argue that our specific design 
should enable an array of boosters as power sources, using a relatively simple architecture for serial biasing relative to RSFQ and efficient variants: energy-efficient RSFQ\cite{KirETAL2011} (ERSFQ) and energy-efficient SFQ\cite{VolETAL2013} (eSFQ). The classical energy barriers that impede transmission and quantum interference in an A-C ring with long Josephson junctions were not reported on previously \cite{Wees1990}, and thus we take the opportunity to discuss this topic as part of the present study.

This article is organized as follows:
In Sec.~\ref{sec:boosterprinciple} we describe the general operation principle
and design challenges for the booster, 
while Sec.~\ref{sec:design} introduces specific designs, describes their 
dynamics and analyzes their performance.
A comparison with boosting of (fixed-polarity) fluxons in other structures 
is presented in Sec.~\ref{sec:comparison_LJJ},
followed by a discussion section ~\ref{sec:discussion},
and a summary of our findings in Sec.~\ref{sec:conclusion}.
Additional analyses are placed in appendices, including extended perturbation analysis of the fluxon boost in Sec.~\ref{app:boost_regularLJJ}.

\section{General booster operation}\label{sec:boosterprinciple}

The design principles of the booster follow from the science of LJJs and SFQ of two polarities. 
RFL gates use LJJs, and in the input and output LJJs of the gates the SFQs, which we here call fluxons,
constitute instances of dispersionless solitons of sine-Gordon type \cite{BaronePaterno, BaroneETAL1971}. 
A Josephson junction (JJ) is considered to be 'long' (and a LJJ) 
if one dimension perpendicular to the stacking direction
is larger than the typical fluxon width,
which is a couple times larger than the Josephson penetration depth $\lambda_J$. When the third dimension is relatively small, the LJJ can be modeled in one spatial dimension,
cf.~Fig.~\ref{fig:boosteroperations}.
Because of the invariance of the sine-Gordon equation under a Lorentz boost,
a sine-Gordon soliton is characterized by the energy-momentum relation of a relativistic particle,
and changing (especially raising) the fluxon velocity is usually called boosting \cite{SGModel_and_applications}.
As alluded to above, a dc-bias through an LJJ will exert a force on a fluxon, where the force direction is determined by the relative sign of the current direction and fluxon polarity such that a given dc-bias current accelerates (boosts) a fluxon, but decelerates an antifluxon. 
In our RFL logic, the bit states are represented by the two fluxon polarities,
cf.~Fig.~\ref{fig:boosteroperations}; note that our mapping between polarity and bit state is updated to the RSFQ convention 
since we now assign the bit state `1' to a fluxon with a clockwise(CW)-directed current vortex.
In our logic, a constant current bias applied to a regular LJJ is inadequate to boost bits 
because one of the bit states will gain energy while the other state loses energy.
The proposed booster solves this problem and allows one to accelerate slow fluxons of either polarity to a high constant velocity, as needed for a sequence of ballistic RFL gates. 

The operation principle of the fluxon booster is illustrated in Fig.~\ref{fig:boosteroperations}. 
First, consider the topology of the device without the bias wiring. 
It is equivalent to an A-C structure made of LJJs: 
it consists of an input and an output LJJ, each connected in series to the two arms of an LJJ ring. 
Thus, the connections form so-called S-branches~\cite{NakajimaETAL1976, NakajimaETAL1978}. 
An S-branch contrasts the common T-branch used as power dividers which, in contrast, connects an input to two ports in parallel~\cite{WilkinsonPowerSplitter}.
For reference, let us first discuss the dynamics in an unpowered S-branch:
in this situation, a fluxon coming from one of the three LJJs will penetrate evanescently into both arms of the LJJ ring. 
However, classically it can not enter both of them, 
since this would require the local creation of a second fluxon, 
which could only occur through the local creation of a fluxon-antifluxon pair. 
At the moderate (slightly relativistic) fluxon energies that are relevant for RFL, this is energetically forbidden. 
Classically, it can also not enter one of the arms exclusively 
since they are equivalent and neither provides an energetic advantage over the other. 
Therefore, a typical S-branch
presents a potential barrier at which a fluxon reflects back into the input LJJ 
(a fluxon with a small input velocity could also become trapped in a shallow potential well formed by a low critical-current point defect before the S-branch).

However, in our fluxon booster, the dynamics at the input branch are modified by dc-bias currents
$I_{\text{b}1} = I_{\text{b}2} = \Ib$, 
which bias the two arms of the A-C ring, as shown in Fig.~\ref{fig:boosteroperations}. 
Crucially, the bias currents on the upper and lower arm
have opposite directions, in contrast to the circuits of
phase mode logic \cite{NakajimaETAL1976, NakajimaETAL1978, YamashitaETAL1995}.
The two opposite bias currents create two (overlapping) potential steps in the ring LJJs, each of the magnitude $\Phi_0 \Ib$. 
This results in a force on a fluxon (antifluxon) that is directed clockwise (counterclockwise). 
These forces act on the front of the fluxon coming in on the input LJJ and thus draw the fluxon into the A-C ring. 
Depending on the force direction (and hence the fluxon polarity), the fluxon will be directed into the upper or the lower arm, as illustrated by panels (a) and (b). 
Inside the LJJ ring, the fluxon is further accelerated near the bias current.
Eventually, the accelerated fluxon reaches the merge point. 
Since this is another S-branch and it is unpowered, 
one could expect the fluxon to stop or back-reflect in the manner described above 
for the case of zero bias current.
However, the dynamics at the merge point differ qualitatively from that of the branch point due to a homogeneous screening current in the rails of the ring caused by the fluxon's presence in the ring. 
Unless the arms of the LJJ ring are very long, $\gg \lambda_J$, 
this screening current is non-negligible, cf.~App.~\ref{app:screeningcurrents}.
As such, if a fluxon is present, 
the resulting screening current in the outer rail of the LJJ ring
acts as an intrinsic bias on the output LJJ. 
As a result, the fluxon will tend to stay in the ring rather than exit on the intended port.

\begin{figure*}
\includegraphics[width=\textwidth]{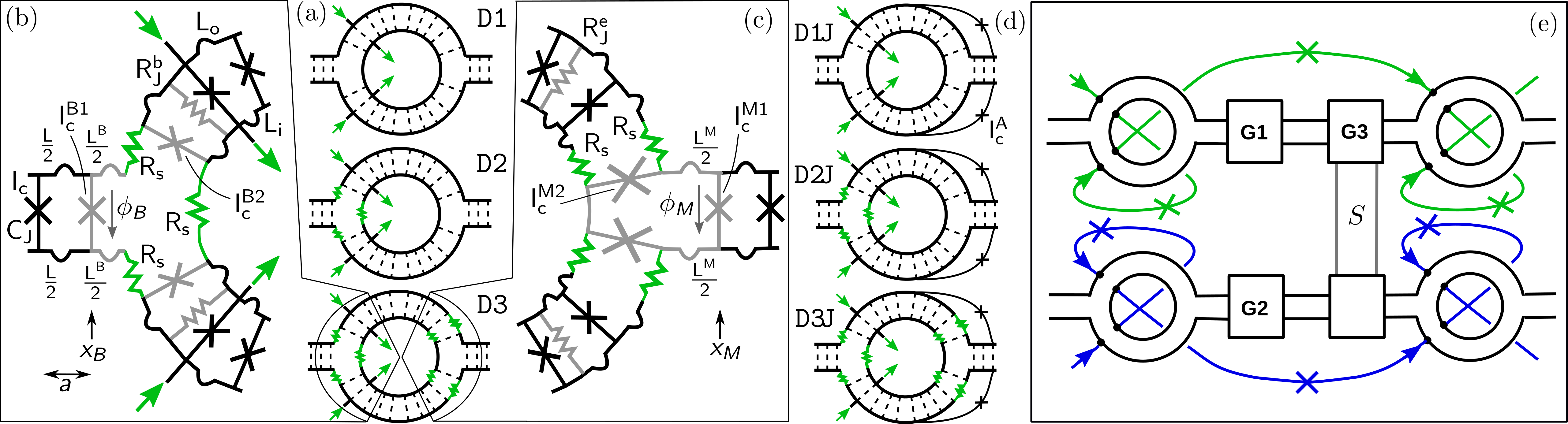}
\caption{
(a,d) Circuit diagrams of the booster, (a) without or (d) with a JJ-bridge,
both with three options for supercurrent isolation: without series resistors ($\Done$, $\DoneJ$), 
with series resistors (green) at the branch cell 
for added isolation between the bias currents ($\Dtwo$, $\DtwoJ$), 
and with series resistors at the branch and merge cells 
for full supercurrent isolation of the bias currents ($\Dthr$, $\DthrJ$).
(b,c) Circuit schematics near the branch cell (the loop including the JJ at position $x_B$ and the two JJs to the right) and merge cell (the loop including the JJ at position $x_M$ and the two JJs to the left. 
Repeating-cell structures are shown in black and unique circuit elements in gray and green. 
The top and bottom arms of the booster are fully symmetric. 
(a-d) Simulations in this work are performed with two current sources per booster (green arrows).
(e) A design proposed for future work with serial biasing of multiple boosters and gates (G1--G3) using $\Dthr$ (see section \ref{sec:design} A). 
Gate G3 contains a stored state $S$ shared by two bitlines (see energy estimates in section \ref{sec:discussion}).
Routes for serial biasing (current recycling) of the boosters are shown in green and blue, with JJs placed in series to add inductive isolation between connected booster arms.
}
\label{fig:boosterschematics}
\end{figure*}

For the fluxon to be released from an LJJ ring into the output LJJ,
modifications near the merge point are required. 
We report on two options:
In the first option, 
the critical currents in the arms of the LJJ ring
are increased near the merge point. 
In the discrete booster circuit, this is enabled by 
increased areas of JJs in the merge cell that are also part of the A-C ring,
cf.~Fig.~\ref{fig:boosterschematics}(c).
For a fluxon approaching the merge point in the upper arm of the LJJ ring, the large JJ on the lower arm essentially acts as a small inductance connection between the inner rail of the ring with the lower rail of the output LJJ, 
and allows the fluxon to enter the output LJJ. 
A short delay in this transmission through the merge cell is caused by the potential barrier of the large JJ in the upper arm. 
In the second option, ``bridge'' connections are made between the outer rails of the ring LJJs and the rails of the output LJJ, 
cf.~Fig.~\ref{fig:boosterschematics}(d).
Each bridge consists of a large JJ and
serves to buffer a part of the boost energy
and to create a temporary current bias in the output LJJ. 
This current bias draws the fluxon into the output LJJ
while transferring the buffered energy back to the fluxon.
Crucially, this is an intrinsic current bias whose direction is determined by 
the fluxon polarity and thus will, in both cases, create a forward-directed force.

\section{Booster design and performance}\label{sec:design} 

Our proposed designs for booster circuits are shown in Fig.~\ref{fig:boosterschematics}. 
Standard LJJs used in this work are composed of discrete cells of total inductance $L$ and an undamped JJ, with critical current $I_c$ and capacitance $C_J$.
The input and output LJJs have balanced rails, so each rail contributes $L/2$ to the total cell inductance. 
In contrast, the LJJs in the arms have cell inductance $L= L_i + L_o$, that are unbalanced and composed of a smaller inner inductance $L_i=0.3L$ and a larger outer inductance $L_o=0.7L$.
In this work the discreteness of all LJJs,
$a/\lambda_J = \sqrt{L/L_J}  = (2\pi L I_c/\Phi_0)^{1/2}$, is equal,
where $a$ corresponds to the cell width, and $\lambda_J$ is the Josephson penetration depth and the characteristic length scale.
The fluxons have a characteristic time scale given by the reciprocal of the plasma frequency $\omega_J = \left( 2\pi I_c/(\Phi_0 C_J)  \right) = 2\pi \nu_J$. 
Since fluxons in our LJJs approximate sine-Gordon solitons,
their energy~\cite{SGModel_and_applications} is 
\begin{equation}\label{eq:Efl}
 \Efl(v) = \frac{8 E_0}{\sqrt{1-(v/c)^2}} 
 \,,
\end{equation}
where $E_0 = \Phi_0 I_c \lambda_J/(2\pi a)$ 
and the velocity is bounded by $|v| < c = \lambda_J \omega_J$.

Our simulations of the schematics use sufficient lengths 
($> 10 \lambda_J$)
of input and output LJJs such that boundary effects are negligible and $L_J/L = \Phi_0 / (2\pi I_c L) = 7$
enables near-continuous LJJ behavior without an impractically large number of cells. 
Note that an LJJ consisting of 40 JJs has a length $\approx 15\lambda_J$.
In contrast, the ring LJJs in our boosters are relatively short, 
using $10$ JJs per ring arm.

While most JJs in the booster circuit follow nominal parameters, 
exceptions are made for JJs in the branch and merge cells, 
where they differ by area, but the ratio of critical current to capacitance, 
$I_c/C_J$, is the same. 
Table \ref{tab:boosterparameters} summarizes these special parameters 
of our boosters.
The upper and lower arms of the ring LJJs are symmetric 
such that our circuit will equally boost either polarity of the input fluxon. 
In each arm, the second JJ from the branch cell is current-biased. 
The biased JJs and a JJ pair near the merge cell 
are shunted with parallel resistors $R_J^{b}$ and $R_J^{e}$, respectively.
The shunt resistors serve to dissipate fluctuations 
generated during the passage of the fluxon through the branch and merge cells. 
In our designs, the added damping is slightly subcritical, 
e.g.~with damping rates
$\alpha_b = R_{J,\text{crit}}/R_J^{b} = 0.4$
and $\alpha_e = R_{J,\text{crit}}/R_J^{e} = 0.2$, 
where $R_{J,\text{crit}} = \sqrt{\Phi_0/(2\pi C_J I_c)}$ 
is the resistance for critical JJ damping. 
However, we find that the booster also works 
with higher damping rates, $\alpha_{b,e} \approx 1$,
with somewhat reduced boost efficiency. 

The two general booster types sketched in panels (a) and (d)
of Fig.~\ref{fig:boosterschematics}, respectively, 
differ in the method by which the fluxon 
is enabled to leave an LJJ ring to the output LJJ at the merge cell: 
The boosters of panel (a) have relatively large $I_c^{M2}$ (using large JJ areas)
in the merge cell,
whereas these JJs are not quite as large in the boosters of panel (d). 
Instead, the latter boosters have an added JJ bridging between 
the outer rails of the ring LJJs and the rails of the output LJJ.
The bridges consist of a large JJ with $I_c^A$
in series with a (small) inductance $L^A$
and connect e.g.~the 6th JJ in each arm of the LJJ ring
with the 2nd JJ of the output LJJ. 

The three booster designs shown in each part (a) and (d)
of Fig.~\ref{fig:boosterschematics} are very similar, 
but differ by their use of small series resistors for added isolation of the bias currents $I_{b1}$ and $I_{b2}$. 
The resistors are either: entirely absent ($\Done$, $\DoneJ$), 
are present in the branch cell for added isolation of the bias currents ($\Dtwo$, $\DtwoJ$),
or are present near the branch and merge cell for full supercurrent isolation of the bias currents ($\Dthr$, $\DthrJ$).
While our circuit simulations show that all designs can boost an input fluxon, 
we argue that $\Dthr$ and $\DthrJ$ are appropriate for serial biasing, as discussed next. 

\subsection{Fluxon booster designs for serial biasing}

It has previously been shown \cite{WusOsb2020_RFL, OsbWus2020_CNOT, OsbWus2023_BSR}, that unpowered (ballistic) RFL gates can operate 
with high efficiency, e.g., the output fluxon energy can be above 
90 \% of the input fluxon value.  
This is very high compared to irreversible gates which usually dissipate 
$\Ib \Phi_0$ per switching of a powered JJ. 
Nevertheless, the gradual energy loss of a fluxon traversing through a sequence 
of RFL gates will eventually cause the ballistic gate operation to fail.
Therefore, we need to periodically include boosters,
which restore the fluxon energy to a nominal value. 
For example, Fig.~\ref{fig:boosterschematics}(e) shows
a $2 \times 2$ gate array with boosters.  
In this diagram, the gates could be BSRs with two input and output ports, 
where input fluxons are powered to the right through BSRs by boosters, 
and the second input ports of the BSRs can also be accessed by fluxons flowing from top to bottom through them. 
A key feature of the booster designs is that they may allow serial biasing (for small total current and input wiring).
If each bias line delivers one bias to an array of $N_{\text{booster}}$ boosters,
where it biases the upper ring arm and the lower ring arm, 
cf.~panel (e),
then the average voltage 
$\langle V \rangle =  N_{\text{booster}} f_{\text{bit}} \Phi_0$ 
builds up along the bias line, 
for bits with frequency $f_{\text{bit}}$. 
This proposed scheme is simpler than state-of-the-art serial biasing for RSFQ and ERSFQ, e.g., Ref.~\onlinecite{CurrentRecycling1}, which uses large islands to route bias current and uses mutual inductance for signals on and off of each island. 

\begin{table*}
\renewcommand\arraystretch{1.2}
\caption{
Circuit parameters
for booster of Fig.~\ref{fig:boosterschematics},
with $10$ JJs per ring arm, $L^{B} = L^{M} = 0.1 L$, 
$I_c^{B2} = 1.1 I_c$, 
and $R_s = 0.0094 \sqrt{L/C_J}$, where $\sqrt{L/C_J}$ is the characteristic impedance of the LJJ (e.g., if $I_c=7.5 \umicroA$, $L/L_J=1/7$, and $C_J$=300fF, then $R_s=0.04 \uOhm$).
The characteristic decay time of power fluctuations in the ring LJJs $\tau_{R} \approx  (\bar \alpha 2\pi \nu_J)^{-1}$,
is determined by the average damping rate 
$\bar \alpha =  \sum_{n=1}^{N=10} \alpha_n = (\alpha_b + \alpha_e)/10$
of the JJs in the ring LJJs, 
with $\alpha_{b,e} = R_{J,\text{crit}}/R_J^{b,e}$
(while series resistors $R_s$ contribute here only 
weakly to the overall damping).
They are determined with the criterion $v_f \geq 0.5 c$ for $v_i = 0.1 c$
by varying a single circuit parameter, while keeping all other parameters fixed.
JJ critical current and capacitance 
are varied separately.  
The parameter margin range includes $-100\%$ to above $+200\%$ 
except if given below.
}
\begin{ruledtabular}
\begin{tabular}{cccc|ccccccc|cc}
& $\displaystyle\frac{\Ib}{I_c}$ 
& $\alpha_b + \alpha_e$
& $\nu_J \tau_{R}$
& $\displaystyle\frac{I_c^{B1}}{I_c}$ 
& $\displaystyle\frac{I_c^{M1}}{I_c}$ 
& $\displaystyle\frac{I_c^{M2}}{I_c}$ 
& $\displaystyle\frac{I_c^A}{I_c}$
& $\displaystyle\frac{L^A}{L}$
& $\displaystyle\frac{L_o}{L}$ 
& $\displaystyle\frac{L_i}{L}$ 
& $\displaystyle\frac{v_f}{c}$
& $\eta$ \\
\hline
$\Dthr$ 
&&& 
& 1.5 & 1.7 & 4.2 & NA & NA & 0.7 & 0.3 &&\\ \hline
&2.8 &0.2 &8.0 
 & $(-100, +48)$\% 
 & 
 & $(-27, +39)$\% 
 && 
 & $(-52, +80)$\% 
 & 
 & 0.68 & 0.37 \\
&2.8 &0.6 &2.7
 & $(-100, +48)$\% 
 & 
 & $(-29, +26)$\%  
 &&
 & $(-51, +62)$\% 
 & 
 & 0.67 & 0.35\\ 
 \hline
\DthrJ
&&&& 1.5 & 1.5  & 2.0 & 10 & 0.5 & 0.7  & 0.3 &&\\ \hline
&2.9 &0.2 &8.0 
 & $(-100, +46)$\%   
 & $(-100, +160)$\%
 & $(-46, +56)$\% 
 & $(-90, \phantom{+100})$\%
 && $(-41, +86)$\% 
 & 
 & 0.69 & 0.38 \\ 
&2.9 &0.6 &2.7
 & $(-100, +46)$\%  
 & $(-100, +96)$\%
 & $(-57, +31)$\% 
 & $(-90, \phantom{+100})$\%
 &
 &$(-36, +56)$\% 
 & 
 & 0.69 & 0.39 
\end{tabular}
\end{ruledtabular}
\label{tab:boosterparameters} 
\end{table*}

For several boosters to share the same two current sources, the dc-bias currents through the device must be at least dc-isolated from one another, 
even in a circuit with slight asymmetries resulting from fabrication uncertainties. 
To that end, we use serial resistors $R_s$ in the rails of the booster 
ring as shown in the designs $\Dtwo$ and $\Dthr$ of Fig.~\ref{fig:boosterschematics}(a), 
and in designs $\DtwoJ$ and $\DthrJ$ of Fig.~\ref{fig:boosterschematics}(d). 
These resistors constrain the dc-bias currents within the A-C arms.
Although the resistors break fluxoid quantization in the booster,
by setting a relatively small value for $R_s$ we ensure that the fluxons transmit through the LJJ ring. 
Our final designs for a booster are $\Dthr$ and $\DthrJ$. 
These have an additional stage of isolation relative to $\Dtwo$ and $\DtwoJ$, resulting in full supercurrent isolation between the ports, and thus should allow serial biasing (routing a current for bias through multiple cells serially).
In principle, the location of series resistors may be chosen differently, 
as long as they isolate the superconducting path between 
the multiple bias current ports. 
However, in our circuit simulations, we find that the booster performance 
(output velocity, booster efficiency, margins) 
is negatively affected when series resistors are used in the merge cell (not shown). For that reason, designs $\Dthr$ and $\DthrJ$ have series resistors in the adjacent ring cells instead. 
As discussed in Sec.~\ref{sec:performance}, the added damping losses from the series resistors cause a small but worthwhile cost to the booster performance.

\subsection{Dynamics}\label{sec:dynamics}

To obtain performance data, 
we have performed numerical simulations of the fluxon dynamics in a circuit containing a single booster with two independent current biases, as shown in Fig.~\ref{fig:boosterschematics}(a) and (d). 
The simulations assume a fluxon initially located at an initial position $X_i$ in the input LJJ moving with small initial velocity $v_i>0$ towards the booster. 
This is done by initializing phases and voltages of the JJs in the input LJJ to the known form of the equilibrium fluxon fields,
$\phi_n = 4\arctan( \exp(-\sigma \theta_n) )$
and 
$V_n = (2\sigma v_i/c)\,     \text{sech}(\theta_n)/\sqrt{1-v_i^2/c^2}$,
at the JJ nodes $n$. 
Herein $\sigma = \pm 1$ is the fluxon polarity 
and $\theta_n = ( x_n - X_i )/(\lambda_J \sqrt{1-v_i^2/c^2})$, 
where the node positions $x_n$ have mutual spacing $a$.
The JJ phases $\phi_n(t)$ at later times $t$ are obtained from the circuit simulation. 
Using the above fluxon shape, 
we fit the concatenated phase distribution $\phi_n(t)$
of the input LJJ, one arm of the ring LJJ, and the output LJJ 
at all times $t$ 
to obtain the fluxon trajectory $X(t)$ 
and its velocity $v=\dot X$. The choice of the ring arm to use in the fitting depends on the arm that the fluxon enters, which depends on the fluxon polarity (the upper arm for $\sigma=1$ and the lower arm for $\sigma=-1$). Although fluxons take a definite path in our structures, at the branch point some kinetic energy from the fluxon is converted into plasma waves that are emitted into both arms of the ring. 

\begin{figure}[t]
\includegraphics[width=\columnwidth]{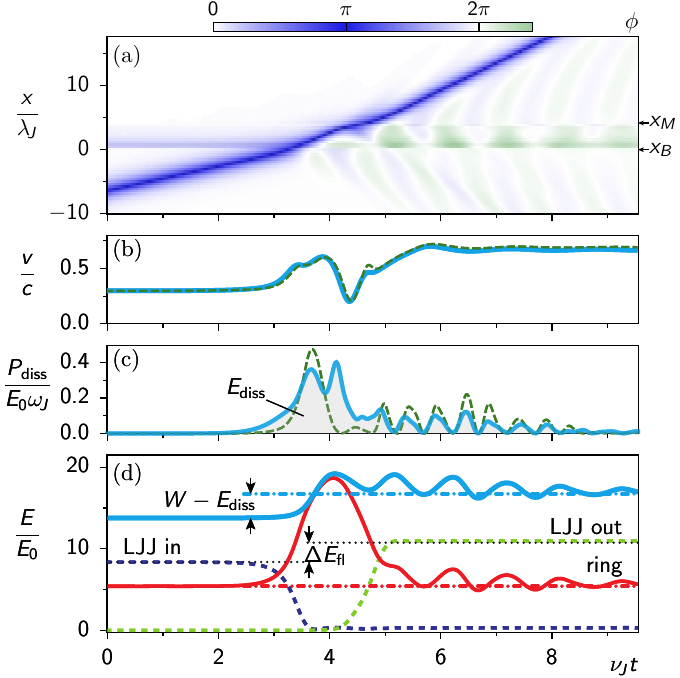}
\caption{
Dynamics of booster $\Dthr$ for $\Ib = 2.8 I_c$, 
and an input fluxon with initial speed $v_i = 0.3c$. 
(a) Dynamics of JJ phases $\phi_n(t)$ in input LJJ (at positions $x_n<x_B=0$), 
upper ring LJJ ($x_B \leq x_n < x_M = 10 a \approx 4 \lambda_J$), 
and output LJJ ($x_n > x_M$). The vertical axis shows position $x$ (with branch and merge points) and the horizontal axis shows time. 
(b) Fluxon speed $v$, determined from fits of $\phi_n(t)$ to a standard fluxon phase distribution.
(c) Dissipated power $P_{\text{diss}} = \sum_{k} I_k V_k$ from all 
resistors $k$ of the circuit;
(d) Energy of the booster circuit (without bias potential) $E^{(0)}$ (blue) and subcircuits:
input LJJ (dark blue dashed), LJJ ring (red solid) 
and output LJJ (light green dashed). 
The dotted black lines indicate the fluxon energies $\Efl(v_i)$ and $\Efl(v_f)$
before and after going through the booster.
(The asymptotic energy of the output LJJ is $2 \%$ larger 
than $\Efl(v_f)$ because plasma waves are generated during the dynamics.)
The dash-dotted lines show final asymptotic energies of the total circuit (light blue)
and LJJ ring subcircuit (red).
The value of $E^{(0)}$ is increased by the work $W=\Ib \Phi_0=6.7 E_0$ 
of the current sources 
minus the dissipation losses $\Ediss = 3.7 E_0$.
The fluxon energy is boosted by $\Delta \Efl = 2.4 E_0$
and the total loss in the boost process is $W - \Delta \Efl = 4.3 E_0 = 0.65 \Ib \Phi_0$. 
The remaining $0.6 E_0$ are carried away as plasma waves.
The data in (a,d) is of design $\Dthr$, 
with two JJs shunted with subcritical-damping resistors
($\alpha_b = R_{J,\text{crit}}/R_J^b = 0.4$ at $n=n_b$ 
and $\alpha_e = R_{J,\text{crit}}/R_J^e = 0.2$ at $n_b+6$). In (b,c) we show the same case (blue solid line); an extra case for $\Done$ with equal total damping rate $\sum_n \alpha_n = 0.6$  applied only at location $n_b$ (green dashed line) shows as expected a single dominant dissipation peak (c). 
} 
\label{fig:dynamics_SBL_Rs}
\end{figure}

Let us first discuss the dynamics of the 
booster design $\Dthr$ shown in Fig.~\ref{fig:boosterschematics}(a). 
The design using JJ-bridges, $\DthrJ$ as shown in Fig.~\ref{fig:boosterschematics}(d), is discussed in App.~\ref{app:Csh3_boosters} because the performance is only slightly improved over design $\Dthr$. 
A typical example of the dynamics is presented 
in Fig.~\ref{fig:dynamics_SBL_Rs}.
Panel (a) shows the JJ phases $\phi_n(t)$
in the input LJJ ($x_n \leq x_B = 0$), 
in the upper arm of the A-C ring 
($x_B < x_n < x_M = (10+1) a \approx 4.2 \lambda_J$), 
and in the output LJJ ($x_n \geq x_M$).
The color map is chosen to highlight the fluxon center, which is seen to move towards the branch point $x_B$ and then undergo acceleration while entering into the upper booster arm. 
When approaching the merge point $x_M$, the fluxon slows down visibly.
This is a consequence of the potential barrier imposed by the increased critical current (and area) of the merge-cell JJs, cf.~Fig.~\ref{fig:boosterschematics}(c).
After a short delay, the fluxon exits into the output LJJ and resumes there at increased speed.
The non-monotonic increase in fluxon velocity is evident from the fluxon speed $v$ shown in panel (b). 
Nevertheless, after temporary slowing, the fluxon gains a stable final velocity in the output LJJ that is much higher than its initial velocity. 

The entry of the fluxon into the ring is affected by screening currents. 
As calculated in App.~\ref{app:screeningcurrents}, 
a fluxon's presence in the A-C ring gives rise to screening currents 
in the rails of the LJJ ring,
and unless the ring LJJs are very long, $\gg \lambda_J$, these 
are non-negligible. 
The screening current on the outer rail 
creates an inherent current bias on the branch JJ (at $x_B$), 
which in turn exerts a force on the entering fluxon. 
This force is always repulsive because 
the screening currents generated by a fluxon are directed CW and those of the antifluxon are directed counter-CW (CCW). 
In the case that the fluxon has fully entered the ring, we can estimate the
screening currents and find that the value of the screening current on the outer rail is proportional to the ratio $L_i/L_o$
between the cell inductances of the inner and outer rail in the A-C ring,
cf.~\Eq{eq:Is_outer}. 
Therefore, we expect the repulsive force on the entering fluxon to be the largest 
if $L_i/L_o > 1$.
In agreement with this expectation for this limit, we observe (not shown)  
that the fluxon is slightly decelerated before the branch cell (unless the 
attractive force created by $\Ib \gg I_c$ overpowers this effect).
After the fluxon is drawn into the ring by $\Ib$, it undergoes a more gradual boost, 
i.e., with less severe slowing down before the merge cell. 
The boosters presented have $L_i/L_o=0.43$, 
cf.~Table \ref{tab:boosterparameters}, and thus fall in the opposite regime of $L_i/L_o < 1$ (which is also geometrically plausible).
Here the fluxon undergoes a faster initial boost, 
but later near the merge cell experiences a more dramatic slowing,
as seen in panel (b).
Nevertheless, we find that in the regime $L_i/L_o<1$ the output speed $v_f$ is
larger than in the opposite regime. 

The branch topologies at the input and output of the A-C ring create considerable perturbations of the fluxon motion, resulting in the excitation of plasma waves. 
These can be seen quite clearly in panel (a). 
In the relatively long input and output LJJs, these waves radiate away from where they are created such that their amplitudes remain small.
Whereas plasma waves emitted into the A-C ring build up to much larger amplitude
due to its short length and closed structure. 
These oscillations are dissipated over time by the resistors in the booster design. 
Panel (c) illustrates the combined dissipation power 
$P_{\text{diss}} = \sum_k I_k V_k$ of all resistors $k$.
The solid blue line refers to the design \texttt{D3} 
(incl. series resistors)
and has two shunt resistors, 
one at the biased JJ at $n_b$ 
and one at $n=n_b + 6$ near the merge cell, with $\alpha_b = 0.4$
and $\alpha_e=0.2$, respectively.
In comparison, the green dashed line refers to design \texttt{D1} (no series resistors) and has a shunt resistor only near the branch cell, 
with $\alpha_b = 0.6$.
During the short time scale of the fluxon passage, the two cases differ qualitatively: the \texttt{D1}-case exhibits a dominant dissipation 
only when the fluxon moves past the biased and shunted JJ. 
In contrast in the \texttt{D3}-case, when the fluxon passes each of the two shunt resistors, that gives rise to the two partly overlapping dissipation peaks. 
Also, the contribution from the series resistors leads to visible broadening 
of $P_{\text{diss}}$, both around the fluxon entry and exit of the ring. 
Design \texttt{D3} seems to have the benefit of
suppressing residual plasma fluctuations slightly more, 
as seen around $\nu_J t \approx 6$.  
Despite these differences on the short time scale, 
both cases show the same average power decay 
$P_{\text{diss}}(t) \propto \exp\left(-\bar \alpha \omega_J t\right)$
when observed over a longer time (not shown), where 
$\bar \alpha = \sum_{n=1}^{N=10} \alpha_n/10 = (\alpha_b + \alpha_e)/10 = 0.06$ 
is the average damping rate in the A-C ring.

\subsection{Performance}\label{sec:performance}

\begin{figure}
\includegraphics[width=\columnwidth]{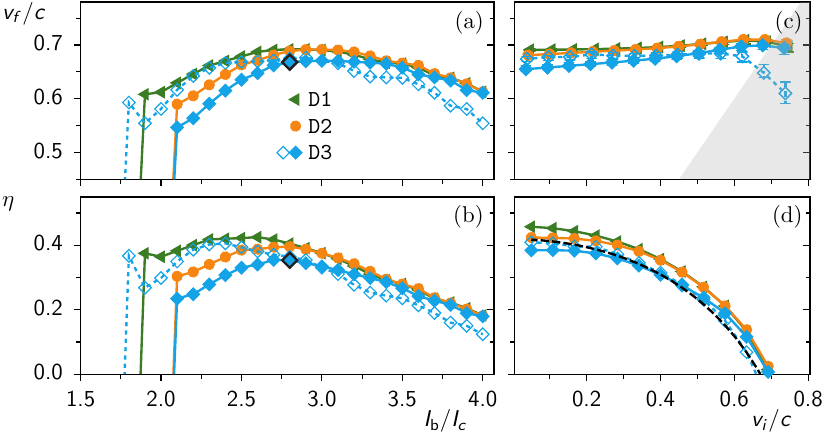}
\caption{
Booster performance: $\Ib$-dependence of 
(a) fluxon output velocity $v_f$ and (b) booster efficiency $\eta$,
and (c,d) their dependence on initial fluxon velocity $v_i$,
for the three booster designs of Fig.~\ref{fig:boosterschematics}(a),
using shunt resistors with
$R_{J,\text{crit}}/R_J^b = 0.4$ and $R_{J,\text{crit}}/R_J^e = 0.2$  (filled markers) 
or 
$R_{J,\text{crit}}/R_J^b = 0.13$ and $R_{J,\text{crit}}/R_J^e = 0.07$ (open markers), respectively. 
In (a,b) the input velocity of the fluxon is set to $v_i = 0.3c$. 
A $v_f$-maximum appears around $\Ib=2.8 I_c$ in the final design $\Dthr$ (emphasized marker) and the margins tabulated in table \ref{tab:boosterparameters} and dynamics illustrated in Fig.~\ref{fig:dynamics_SBL_Rs} refer to this point.
In (c,d) the bias is fixed at that value, $\Ib=2.8 I_c$, 
and we note that $v_f$ is $v_i$-insensitive over a wide $v_i$-range,
before decreasing into the non-boost regime, $v_f \leq v_i$ (shaded area).
The dashed line in (d) shows an estimate for $\eta$ assuming constant $v_f$.
}
\label{fig:boosterperformance}
\end{figure}

From the simulated fluxon dynamics, we obtain performance data
as functions of the bias current $\Ib$ and the initial velocity $v_i$. 
Figures \ref{fig:boosterperformance} and \ref{fig:boosterperformance_Csh3} illustrate the performance of the various
booster designs of Fig.~\ref{fig:boosterschematics}(a) and (d),
respectively.
All of these designs are found to have similar characteristics
and in the following, we concentrate our discussion on 
Fig.~\ref{fig:boosterperformance}.
Panel (a) shows the boosted output
velocity $v_f$ of the fluxon as a function of the bias current $\Ib$.
The booster has a threshold in $\Ib$, below which an incoming fluxon
is not transmitted through the entire booster. 
Below the threshold, the fluxon may become pinned when first reaching the branch cell
(due to nearby resistors $R_J^b$ and $R_s$ and a shallow potential well 
at the branch point), 
or when reaching the biased and damped JJ, 
or otherwise, it can reflect off the merge cell 
and eventually, pin at the biased JJ on its second encounter. 
These dynamic types occur in this order with increasing $\Ib$-value.
Once $\Ib$ exceeds the threshold, 
the boosted velocity $v_f$ first increases with $\Ib$, as one would expect. 
The boosted velocity $v_f$ reaches a maximum in the current range $\Ib/I_c \approx 2.5 - 3.0$ (depending on the design), before starting to lower for larger $\Ib$.
From our simulations, we conclude that this lowering is 
caused by the high potential barrier near the merge cell, 
where the fluxon loses a large fraction of its boosted energy in the form of plasma waves
when impinging with very high speed. 
This lowering is similarly observed in the case of our alternative booster designs 
$\DoneJ - \DthrJ$, cf.~App.~\ref{app:Csh3_boosters}, 
but may be avoided by careful design of the potential energy in the A-C ring. 
We discuss this aspect in more detail in Sec.~\ref{sec:comparison_LJJ}.
Generally speaking, one may say that the decreasing output velocity at large $\Ib$ 
originates from the problem of routing the fluxon through the merge cell into the output LJJ.

The threshold for booster operation typically lies above $\Ib > I_c$,
cf.~Fig.~\ref{fig:boosterperformance}(a). 
This is different than in RSFQ circuits, 
where each critically damped JJ is individually biased with $\Ib \approx 0.7 I_c$.
In our case, however, the circuit forms (weakly) discrete LJJs, 
where the small cell inductance $L$
causes the bias current to spread over several JJs within an arm,
each of which is thus biased sub-critically.

The data shown in Fig.~\ref{fig:boosterperformance}(a)
are obtained using $\alpha_b=0.4$ and $\alpha_e=0.2$ (filled markers)
at $n=n_b$ and $n=n_b + 6$, respectively. 
We note that very similar $\Ib$--$v_f$-curves are obtained 
for other distributions of shunt resistors, 
as long as the sum of damping rates $\sum_n \alpha_n$ is the same.
If, for instance, the entire (shunt) damping rate is concentrated 
at a single JJ, e.g. $\alpha_b=0.6$ and $\alpha_e=0.0$,
visible differences occur only for $\Ib$ around the threshold. 
Whereas the lowering of the total (shunt) damping rate $\sum_n \alpha_n$ 
causes an overall shift of the $\Ib$--$v_f$-curves towards smaller $\Ib$
while leaving the maximum boost velocity invariant.
This is illustrated in Fig.~\ref{fig:boosterperformance} 
for the design $\Dthr$, for which  
a case of weaker damping, $\sum_n \alpha_n=0.2$ (open markers, 
$\alpha_b=0.13$ and $\alpha_e=0.07$) is also included.
Thus, as expected, the $\Ib$-threshold increases with the 
total shunt damping rate. 
Similarly, it increases when small serial resistors $R_s$ are added in the branch cells, as seen in the comparison of design $\Done$ with the designs $\Dtwo$, $\Dthr$ for equal shunt resistances (filled markers).
The addition of more serial resistors in the sequence $\Done$, $\Dtwo$, and $\Dthr$ then also reduces the boosted velocity $v_f$ for moderate $\Ib$-values, whereas the position of the $v_f$-maximum and the $v_f$-decline at very large $\Ib$ are similar for all three designs.

The formation of a $v_f$-maximum at moderate $\Ib$ is one of the reasons 
to keep the booster arms relatively short. 
In Fig.~\ref{fig:boosterperformance} and Table \ref{tab:boosterparameters} 
we have chosen $10$ JJs in each arm of the booster ring, 
and we note that the arms then form ``short LJJs'', i.e. 
they are longer than but comparable to $\lambda_J$.
Making the arms even shorter appears to be detrimental 
to the performance at large initial velocities: 
in our simulations, we observe a rapid decrease of $v_f$ as a function of 
$v_i$, for $v_i \gtrsim 0.5$. 
However, for $N=10$, as illustrated in Fig.~\ref{fig:boosterperformance}(c), 
the output velocity of the booster is relatively uniform 
up till the break-even point $v_f \approx v_i$ (in contrast, the no-boost regime $v_f<v_i$ is marked by the shaded area).
This covers the entire range of relevant input velocities $v_i < 0.6c$, 
given that the nominal operation velocity in most RFL gates is set to $v\approx0.6 c$.
Making the booster arms longer ($N>10$), allows the boosted fluxon velocity 
to saturate before exiting the ring, but at a larger value of $\Ib$ 
where the booster efficiency is much reduced (see below).

We calculate the dissipated energy 
$\Ediss = \int \diff t P_{\text{diss}}$
from the dissipation power, cf.~Fig.~\ref{fig:dynamics_SBL_Rs}(c), 
by integrating over a time scale much larger than the characteristic
damping time $\tau_R = 1/(\bar \alpha \omega_J)$ (not shown). 
Monitoring also the circuit energy 
$E^{(0)} =  \sum_n \left[ \Phi_0^2 C_{J,n} \dot \phi_n^2/(8\pi^2) - I_{c,n} \cos(\phi_n) \right] + \sum_i L_i I_i^2/2$
over this time,
see Fig.~\ref{fig:dynamics_SBL_Rs}(d),  
we can infer the external work done by the current source, 
$W = E^{(0)}(t \gg t_{\text{boost}} + \tau_R) - E^{(0)}(0) + \Ediss$.
In cases of a successful fluxon boost, 
where one of the two biased JJs undergoes a $2\pi$-phase change 
during the fluxon's passage, 
this numerically determined value is consistent 
with the formula $W =\Ib \Phi_0$, 
describing the change in potential energy of a single JJ 
through a current bias $\Ib$.
(Using sine-Gordon perturbation theory \cite{McLaughlinScott1978},
we also find $\Ib \Phi_0$ to be the height of the potential energy step 
experienced by a fluxon in a continuous LJJ with a local current bias, 
cf.~App.~\ref{app:boost_regularLJJ}.)
Depending on the value of $\bar \alpha$, the actual dissipated energy makes
up a fraction of this, $\Ediss < W$. 
In comparison, 
the critically damped JJs of RSFQ gate routinely consume 
$\Ediss \approx W = \Ib \Phi_0$.
Moreover, in the proposed architecture of RFL, 
the SFQ will receive only occasional boosts 
and thus much fewer biases are required per logic gate sequence than in RSFQ. 

We calculate the booster efficiency $\eta = \Delta \Efl / W$
as the ratio of the fluxon's energy gain $\Delta \Efl = \Efl(v_f) - \Efl(v_i)$ 
to the external work $W$.
The booster efficiency $\eta$ is shown in Fig.~\ref{fig:boosterperformance}(b)
as function of $\Ib$. 
Over a range of moderate $\Ib$, $\eta$ increases together with $v_f$ (and with $\Delta \Efl$). 
However, $\eta$ assumes a maximum when $v_f$ and $\Delta \Efl$ no longer 
substantially grow with $\Ib$ and upon further increase it decreases because of the $\Ib$-proportional scaling of $W$. 
 
As a function of the initial velocity $v_i$,
the efficiency $\eta$ decreases, as seen in Fig.~\ref{fig:boosterperformance}(d). 
This follows since $W$ is independent of $v_i$
while the $v_i$-insensitive output velocity $v_f$ implies 
that $\Delta \Efl$ is largest for small $v_i$. 
The black dashed line shows an approximation, 
using $\Delta \Efl = \Efl(\bar v_f) - \Efl(v_i)$
with the mean output velocity 
$\bar v_f = 0.67$ of $\Dthr$ in the range $v_i/c \leq 0.6$.
For the data shown in panel (b), which is for $v_i=0.3$, 
and at the operation point $\Ib/I_c = 2.8$, 
the output velocity is $v_f = 0.67 c$
and for the total damping rate $\alpha_b + \alpha_e = 0.6$,
the efficiency is $\eta = 0.35$
($\Delta \Efl = 2.4 E_0$,  $W = 6.7 E_0$, $\Ediss = 3.7 E_0$).
The total energy loss in the boost process is then
$W - \Delta \Efl = 4.3 E_0 = 0.65 W$, 
including the energy lost to plasma waves and the energy dissipated in the resistors.
Since the booster is an intended power source for RFL, 
the added fluxon energy $\Delta \Efl$, the energy dissipation $\Ediss$,
and the total energy loss $W-\Delta \Efl$
should be compared with the work $W = \Ib \Phi_0$, 
which yields the ratios of $35\%$, $56\%$, and $65\%$, respectively. 
Recall also that the energy of a fluxon at nominal speed ($v=0.6c$) is $\Efl = 10 E_0$. 
For the smaller damping rate $\alpha_b + \alpha_e = 0.2$ 
the boost process has slightly improved efficiency $\eta = 0.37$ 
($\Delta \Efl = 2.5 E_0$,  $W = 6.7 E_0$, $\Ediss = 3.4 E_0$, 
$W - \Delta \Efl = 4.1 E_0$).

The booster performance can be compared 
with the performance of the (polarity-dependent) fluxon 
boost in a regular (discrete) LJJ, cf.~Sec.~\ref{sec:comparison_LJJ}.
Using the same values of $a/\lambda_J$, $R_J^b$, $\Ib = 2.8 I_c$, 
and $v_i=0.3 c$  we find that a fluxon in a biased LJJ is boosted to 
$v_f=0.78 c$ with $\eta = 0.64$
($\Delta \Efl = 4.3 E_0$,  $W = 6.7 E_0$, $\Ediss = 1.7 E_0$). 
Compared with this,
the efficiency $\eta = \Delta \Efl/W$ 
in our final booster designs, $\Dthr$ and $\DthrJ$, is reduced by 
a factor of $\approx 1/2$,
owing to the dynamics at the branch and merge cells, 
the boundary conditions imposed by these cells during the fluxon boost, 
and the added series resistors.


\section{Comparison with boost performance in other structures}
\label{sec:comparison_LJJ}

\begin{figure}[tb]
\includegraphics[width=\columnwidth]{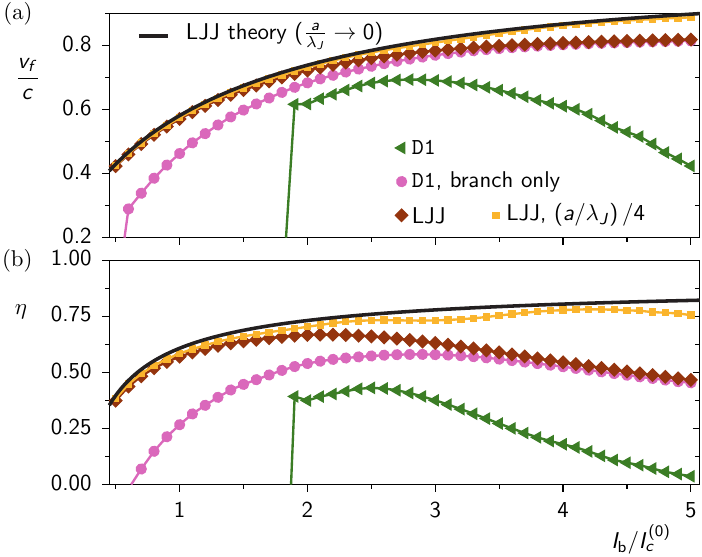}
\caption{
Boost performance in different systems.
$\Ib$-dependence of (a) fluxon output velocity $v_f$ 
and (b) boost efficiency $\eta$
for (i) the ring booster of design \texttt{D1} (green triangles), 
for (ii) a S-branch alone (purple circles), 
and for (iii) an LJJ with discreteness 
$a^2/\lambda_J^2 = 1/7$ (brown diamonds), 
and (iv) with discreteness $\left(a^2/\lambda_J^2\right)' = 1/112 = 1/(7\cdot 16)$ 
(yellow squares). 
The ring booster and the S-branch also have discreteness 
$a^2/\lambda_J^2 = 1/7$.
All simulations use $v_i/c=0.3$ and the same shunt resistance $R_J^b$. 
For the LJJ case with lower discreteness (smaller $L/L_J$), we have assumed that
$(I_c', C_J', L', 1/R_{J,\text{crit}}') = (I_c, C_J, L, 1/R_{J,\text{crit}})/4$
where $R_{J,\text{crit}}$, $R_{J,\text{crit}}'$ 
are the resistances for critical JJ damping. 
In the standard JJ cases (i-iii) the resistance ratio is
$R_{J,\text{crit}}/R_J^b = 0.6$, 
and thus $R_{J,\text{crit}}'/R_J^b = 2.4$ in case (iv).
The black line illustrates the analytic SGP result,
\Eqs{eq:uf_Ib_RJb} and \eqref{eq:eta_SGPA}, which assumes a homogeneous LJJ,
corresponding to the limit $a/\lambda_J \to 0$,
and gives good agreement with the small-discreteness LJJ.
}
\label{fig:boostperformances}
\end{figure}

The booster circuits allow one to accelerate fluxons irrespective of their polarity.
In contrast, to power fluxons of a fixed polarity 
a bias current applied in a single LJJ would suffice, and it would provide a boost with higher efficiency. 
This is illustrated in Fig.~\ref{fig:boostperformances}, 
where we compare the boost performance in different LJJ geometries: 
(i) in the ring booster of design \texttt{D1} (green triangles), 
(ii) a biased S-branch (purple circles) designed in the same way 
as the left part of the $\Done$-booster (i.e. same branch cell and $\Ib$-biased, $R_J^b$-shunted JJs),
and (iii, iv) an LJJ with a single $\Ib$-biased, $R_J^b$-shunted JJ (brown diamonds, yellow squares).
We compare these simulation results 
with the analytic formula, \Eq{eq:uf_Ib_RJb} (black solid line), 
obtained from sine-Gordon perturbation (SGP) analysis
which is detailed in App.~\ref{app:boost_regularLJJ}.
Here and in all of the simulated cases (i-iv) the same value of $R_J^b$ is used, whereas in case of larger $R_J^b$ (not shown) \Eq{eq:uf_Ib_RJb} produces $v_f$- and $\eta$-curves that are shifted further upwards.

The effective discreteness in all LJJ sections of cases (i-iii) is
$a^2/\lambda_J^2 = 1/7$,
whereas case (iv) has smaller discreteness,
$\left(a^2/\lambda_J^2\right)' = 1/112 = \left(a^2/\lambda_J^2\right)/16$. 
The latter case (iv) of small LJJ discreteness lies much closer 
to the SGP approximation,
as is expected since the SGP analysis assumes a homogeneous LJJ, i.e.~the limit $a/\lambda_J \to 0$.
In comparison, in the LJJ (iii) with larger discreteness (brown)
the boosted velocity is reduced at large $\Ib$-values where the boosted 
fluxon velocity $v_f$ is high. 
This can be understood from the additional damping mechanism
arising through the discreteness-induced excitation of plasma waves 
by a fast moving fluxon \cite{WusOsb2020_RFL, BraKiv1998}. 
Examples of the fluxon boost dynamics in the discrete LJJ are shown
in Fig.~\ref{fig:LJJ_fluxonboost},
and the damping of the fluxon after boost to high velocity $v_f$
can be observed in panel (c).
(In this regime of very large $v_f$, 
where discreteness-induced damping is non-neglible,
the value of $v_f$ depends on where it is measured.
We have taken care to measure all cases at the same distance from the 
biased JJ, or from the merge cell in case of the booster.)
At large $\Ib$-values, another effect increasingly plays a role  
for the fluxon boost in an LJJ.
As calculated in App.~\ref{app:boost_regularLJJ}, 
the bias current generates a localized mode
$\phi_{\text{loc}} = A e^{\mu |x-x_b|/\lambda_J}$ 
of finite width $\mu \lesssim 1$ and with amplitude $A \propto \Ib$.
For large mode amplitude $A$, 
the interaction with this mode causes the fluxon to undergo 
two distinct boost phases, respectively caused by the attractive (repulsive) force 
before (after) the mode center. 
The fluxon then typically slows down temporarily 
between these two boost phases, causing a short time delay after crossing 
the bias point, as seen in Fig.~\ref{fig:LJJ_fluxonboost}(b,c).
This effect is independent of $a/\lambda_J$, 
but in combination with the larger damping at finite $a/\lambda_J$,
it prevents the fluxon from reaching the large velocity observed 
in the SGP dynamics (where the latter ignores both the finite discreteness 
and the presence of a localized state). 
In Fig.~\ref{fig:boostperformances}(a) this leads to a $v_f$-saturation trend  
at large $\Ib$ in the (iii) LJJ case (brown diamonds). 
This trend is similarly observed in case (ii) of the S-branch (purple), 
which converges towards the $v_f$-curve of the LJJ.  
At low $\Ib$, $v_f$ in the S-branch is reduced relative to the LJJ
because the branch geometry acts as a perturbation to the regular fluxon shape and motion.

In comparison with the S-branch, 
whose boost threshold lies at $\Ib/I_c \approx 0.6$,
the (i) ring booster (green) 
has a much higher $\Ib$-threshold of operation, $\Ib/I_c \approx 1.9$. 
To a small extent,  
this threshold shift can be attributed to the energy cost incurred from the 
generation of screening currents in the A-C ring mentioned above, 
cf.~App.~\ref{app:screeningcurrents}. 
This energy cost and the resulting small potential barrier at the interface 
of input LJJ and A-C ring 
prevents the fluxon from entering the ring for $\Ib/I_c < 0.8$.
But even when the fluxon can enter the ring, 
the effect of the (damping) losses inside the ring 
as well as the high potential barrier at the merge cell means that the 
boosted energy does not suffice for the fluxon to exit the ring
in the bias range $\Ib/I_c \approx 0.8 - 1.8$.

At large $\Ib$-values, Fig.~\ref{fig:boostperformances}(a) shows 
a steady decline of the output velocities $v_f$ for the ring booster, 
in contrast to all the other structures lacking a closed topology.
This under-performance at large $\Ib$ can be predominantly attributed to the
difficulties in routing the fluxon into the output LJJ. 
As mentioned above, routing of the fluxon from an arm of the A-C ring 
into the output LJJ does not occur for uniform LJJ parameters, 
and therefore this requires special design choices.
In our booster design $\Done$, where it is achieved by very large $I_c^{M2}$-value of the merge cell JJs, the merge cell presents a high, relatively sharp 
potential barrier to the boosted incident fluxon. 
At large $\Ib$, where it initially receives a large boost and impinges on the potential barrier at large speed, a large fraction of the boosted fluxon energy
is turned into plasma waves, thus limiting the output velocity. 
 
Before further exploring the role of the potential in the A-C ring, let us briefly discuss a finite-size effect that limits the performance at large $\Ib$ in a short LJJ ring. 
As mentioned earlier, at large $\Ib$ the fluxon may temporarily slow down during the boost due to 
its interaction with the localized, $\Ib$-generated mode. 
We observed that if the arms of the A-C ring are relatively short and the fluxon velocity is already high after the first boost phase (due to large $\Ib$), the fluxon will reach the end of the arm already during this short time delay. 
As a result, it misses the 2nd boost phase 
and exits the ring with a comparably low output velocity.
If, on the other hand, the booster arms are 
sufficiently long to allow saturation of the fluxon velocity inside the ring, we find that this happens at larger $\Ib$ where the efficiency $\eta$ is already 
much reduced (cf. earlier remark).

\begin{figure}[tb]
\includegraphics[width=\columnwidth]{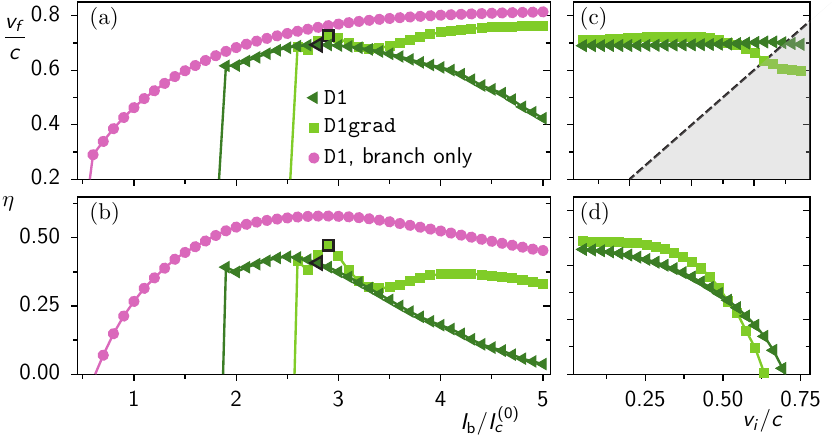}
\caption{
Boost performance in different systems,
with panels equivalent to those of Fig.~\ref{fig:boosterperformance},
here for (i)  the ring booster of design \texttt{D1} (green triangles), 
for (ii) the S-branch shown in Fig.~\ref{fig:boostperformances} (purple circles), 
and for (iii) an alternative booster design \texttt{D1grad} (lime-green squares). 
The JJs within the booster ring of the latter vary linearly in size, 
with $I_c^{B2} = I_c$ at $n=1$ (at the branch cell) 
to $I_c^{M2} = 3 I_c$ at $n=10$ (at the merge cell),
with correspondingly increased values of $C_J$ in the ring, 
whereas all other parameters are those of $\Done$.
All cases use the same value of $R_J^b$ as in Fig.~\ref{fig:boostperformances}
(which in case (iii) implies $R_{J,\text{crit}}'/R_J^b = 0.6/1.22 = 0.49$
because the biased JJ is by a factor $1.22$ larger than a standard JJ).
In case of {\tt D1grad},
the upward potential slope created by the gradually increasing JJ size 
limits the velocity increase of the fluxon in the ring.
As a result, its exit from the ring is a smoother process compared with $\Done$, 
avoiding some of the backscattering at the sharp potential barrier 
of the $\Done$-merge cell.
At very large $\Ib > 4 I_c$ 
the performance thus approaches that of the S-branch, indicating that 
the merging losses relative to the fluxon boost $\Delta \Efl$ are minimized.
Panels (c,d) show dependence on the input velocity $v_i$
at the operation points $\Ib=2.9$ ($2.8$) of the booster {\tt D1grad} ($\Done$), 
defined by a local maximum of their output velocity $v_f$ (for $v_i=0.3 c$) in (a). 
Here, {\tt D1grad} exhibits strongly declining $v_f$ 
much before the break-even point $v_f\approx v_i$, 
in contrast to the $v_i$-insensitive $\Done$,
and for that reason we deem $\Done$ the superior booster despite its somewhat 
smaller $v_f$ and efficiency $\eta$.
}
\label{fig:boosterperformance_modIJac}
\end{figure}

To explore the role of the merge cell potential barrier in some more detail, 
we compare in Fig.~\ref{fig:boosterperformance_modIJac}
the regular booster design $\Done$ (green triangles) with yet another variant {\tt D1grad} (lime-green squares),
where the JJs within the ring vary linearly in size, 
with $I_c^{B2} = I_c$ at $n=1$ (at the branch cell) 
to $I_c^{M2} = 3 I_c$ at $n=10$ (at the merge cell). 
(The JJs at the input and output side of the branch and merge cell, respectively, 
are the same as for $\Done$.)
In contrast to $\Done$, where the sudden increase from $I_c$ at $n=9$ to $I_c^{M2} = 4.2 I_c$ at $n=10$
creates a localized, large-amplitude force against the fluxon motion, 
this force is less localized and of smaller amplitude
for {\tt D1grad} with its the incremental $I_c$-increase.
Despite the reduced value of $I_c^{M2}$ and therefore reduced height 
of the potential barrier, panel (a) shows 
that the $\Ib$-threshold is here shifted to a higher value 
(in the presence of the force from the $I_c$-gradient, 
larger $\Ib$ is required to draw the fluxon into the ring and to overcome the damping at $n_b$).
The force from the $I_c$-gradient also limits the velocity increase within the ring
such that the fluxon receives its largest boost only once it exits 
through the merge cell. 
At very large $\Ib > 4 I_c$,
the performance approaches that of the S-branch (purple circles) and thus indicates that 
the merging losses relative to the fluxon boost $\Delta \Efl$ are minimized.
(In absolute terms, the energy that is either converted into plasma waves 
or is dissipated in resistors, grows with $\Ib$ at large $\Ib$. 
In contrast, $\Delta \Efl$ converges ({\tt D1grad}) or even declines ($\Done$)
with $\Ib$.)
In comparison, in $\Done$ with its comparably sharp merge cell potential,
a much larger fraction of the boosted fluxon energy is turned into plasma oscillations,
leading to the decline of $v_f$ at large $\Ib$.
At $\Ib = 5 I_c$, for instance,
we find that $\Done$ has an output fluxon energy of $\Delta \Efl = 0.4 E_0$,
compared with approximately $E_{\text{pl}} = 9.2 E_0$ of the boost and initial fluxon energy going into plasma waves.
(Of $E_{\text{pl}}$ has contributions from plasma waves which radiate into the input and output LJJ ($1.9 E_0$) and which are dissipated through damping in the ring
($7.3 E_0$), whereas the energy dissipated through damping during the fluxon passage of the ring is $2.1 E_0$).
At the same $\Ib$-value,
{\tt D1grad} has $\Delta \Efl = 3.9 E_0$
compared with $E_{\text{pl}} = 6.3 E_0$ converted into plasma waves
($1.1 E_0$ radiated away and $5.2 E_0$ dissipated through damping, whereas the dissipation during fluxon passage of the ring is $1.5 E_0$).
The combined $\Delta \Efl + E_{\text{pl}}$ is approximately equal in both cases.
These numbers therefore showcase that at large $\Ib$  $\Done$ converts 
a much larger fraction of the boost energy into plasma waves ($95 \%$ at large 
$\Ib = 5I_c$) than the smoother potential design {\tt D1grad} ($\approx 60 \%$).
However, as panel (b) illustrates, 
the boost efficiency at such large $\Ib$ is reduced at such large $\Ib$ 
even for {\tt D1grad}. 

At intermediate $\Ib$, both $\Done$ and {\tt D1grad} have  
a local $v_f$-maximum, see panel (a), 
which is however rather narrow in case of {\tt D1grad} 
and thus requires more careful tuning. 
Furthermore, in case of {\tt D1grad} this maximum is seen in panel (c) to 
strongly decline at large $v_i$,
much before the break-even point $v_f \approx v_i$. 
In contrast, the boost velocity of $\Done$ is nearly $v_i$-insensitive.
That is why, although {\tt D1grad}
offers slightly higher output velocity $v_f$ and efficiency $\eta$, 
we nevertheless deem the booster design $\Done$ as superior in performance.

\section{Discussion}\label{sec:discussion}

We have found that the booster is one device that can asynchronously produce a high-speed fluxon 
from a low-speed fluxon, as needed for RFL. 
Once boosted to high speed, the fluxon is meant to run through a number $k \geq 2$ 
of sequentially arranged ballistic gates before encountering the next booster.
The number $k$ will depend on the involved gate types:
while the ballistic shift register \cite{OsbWus2023_BSR} allows for $k=2$,
we anticipate that the fundamental 1-bit RFL gates \cite{WusOsb2020_RFL} 
allow for $k>2$ because of their wider velocity margins.
The work performed in boosting a bit is $\Ib\Phi_0$, and does not depend on the data-input state. 
However, the total energy lost per boost is smaller, $\approx 0.65 \Ib\Phi_0$, only some of which is dissipation.
On the other hand, RSFQ and energy-efficient RSFQ variants (including ERSFQ, and eSFQ) have a higher dissipation of $\Ib\Phi_0$ per JJ switching \cite{Mukhanov2011}.

To analyze efficiency, we first discuss a standard eSFQ gate, the D flip-flop used in a shift register \cite{VolETAL2013}. 
This gate (see Fig.~5a of Ref.~\onlinecite{VolETAL2013}) switches 3 or 4 JJs from the clock SFQ and data bit, depending on the data bit state, 
and thus the average energy dissipation is $3.5 \Ib \Phi_0$ per data bit. 
We compare this with an RFL circuit as shown in Fig.~\ref{fig:boosterschematics}(e), where a (horizontal) sequence of $k=2$ RFL shift registers along two bitlines are powered by one booster per bitline, 
with an energy loss of $0.65 \Ib\Phi_0$ per bitline.
In this circuit, gates G1 and G2 perform standard 1-bit shift-register operations, while gate G3 is a two-input shift register, which may perform a switch operation.
After data input to the 2-input shift register on one of the bitlines, the same bitline carries the stored bit state $S$ as output. 
However, if the bitlines carry data bits in alternating order, the data switches from one bitline to the other. 
In this case, the energy loss per data bit and logical depth $k$ would be $\approx 0.33 \Ib\Phi_0$. This comparison reveals a $3.5/0.33=10.6$ lower energy loss in RFL compared to eSFQ, and we estimate that RFL may give an order of magnitude improvement in energy efficiency over eSFQ.

We used an A-C ring in our circuit because it allowed us to bias both fluxon polarities separately in a classical fluxon regime. As we have found in our study, the entry of a classical fluxon into the
LJJ ring requires bias currents, and its exit at the merge point requires
significant modifications to the LJJ ring parameters.
Thus, an unpowered LJJ ring (without bias currents)
simply connected to identical LJJ ports, will not allow fluxon transmission classically.
One may reasonably wonder if the potential barriers at entry and exit would also affect the transmission in the quantum regime, at least in principle, 
where one expects to observe the A-C effect as the self-interference of a fluxon transmitting through both ring arms at the same time. 
We have not seen this question addressed in the A-C effect literature (some subset of which uses the term Aharonov-Bohm) \cite{Wees1990}. Additionally, past experiments only explore the A-C effect indirectly through a change in resistance perpendicular to the direction of fluxon flow \cite{EliETAL1993}. It is known that this circuit theoretically allows a fluxon to enter and exit the ring if the barrier(s) are low enough. Our study describes the nature of two classical barriers qualitatively. Thus in principle, an extension of this study may enable a more direct understanding of the A-C effect, i.e., the quantum tunneling of a fluxon through an A-C ring structure with LJJs on the ports.

\section{Conclusion}\label{sec:conclusion}

Here we simulate power sources, named boosters, for our ballistic logic family named RFL. 
The RFL family includes BSRs, which exhibited high energy efficiency and asynchronous clocking in previous simulations. 
In the BSR and other gates from the same logic family, the bit states are unconventional degenerate states but the power source for operating a sequence of gates was not specified. 
Here we describe boosters, based on A-C rings, that may power the RFL BSRs and other ballistic gates. 
According to simulations, the boosters increase the fluxon velocity as needed. 

In the simulations, we used an independent current bias per arm of the booster. 
Additionally, booster designs $\Done$--$\Dthr$ have different levels of bias current isolation between booster arms. 
Design $\Dthr$ has full supercurrent isolation between the arms and relates to serial biasing, a possible method for scaling SFQ logic. 
Our proposed method of serial biasing is relatively simple because it involves no mutual inductors; we save a study of our serial biasing for future work. 

We explain the operation and performance of the boosters related to two fluxon polarity states, fluxon dynamics, JJ modifications, and supercurrent-isolating resistors in three different designs ($\Done$--$\Dthr$). 
The data show uniformly high output velocity with good efficiency, reasonable margins, and simplicity of design. 
A challenge found for the design of these boosters is related to the fluxon dynamics near the merge cell, where the arms of the ring connect to the output LJJ. 
To allow the exiting of the fluxon from the A-C ring, large critical-current JJs are used in the merge cell. 
Alternative designs with added JJ bridges ($\DoneJ$--$\DthrJ$) allowed the fluxon to transmit through this cell into the output LJJ with a more monotonic acceleration, but the total performance was similar to the base designs ($\Done$--$\Dthr$). 
In this study, we also compared the booster performance with the fluxon boost in a reference circuit that is simply a biased LJJ, a case where the dynamics can be analyzed using fluxon perturbation theory.

In comparison with standard RSFQ logic, where an SFQ passes through many biased JJs per gate, 
the boosters are efficient within the proposed register architecture because the SFQ bit only needs to pass through one biased booster per ballistic gate sequence. 
Using a comparison with two types of RFL shift registers and a RSFQ-variant shift register circuit, we estimate that RFL is an order of magnitude more efficient. 
In the future, we look forward to demonstrating the booster as a power source for ballistic logic circuits.

\FloatBarrier
\begin{acknowledgments}
KDO is grateful for scientific discussions on digital logic with Q. Herr, A. Herr, V. Semenov, M. Frank, I. Sutherland, B. Sarabi, G. Herrera, and N. Yoshikawa, and also broader scientific discussions with C. Richardson, B. Butera, B. Palmer, F. Gaitan, R. Lewis, and A. Murphy. KDO and WW would like to thank H. Cai for recent discussions on experiments. WW would like to thank the University of Otago Physics Department for hosting her.
\end{acknowledgments}

\begin{appendix}

\section{Screening currents}\label{app:screeningcurrents}

Owing to flux quantization in a superconducting loop, 
the entry of a fluxon into the A-C ring gives rise to screening currents in 
the rails of the LJJs arms. 
To estimate the screening current, we consider a fluxon sitting 
in the LJJ ring where it contributes a current $|I_{\text{fl},n}|$
on the rails in each cell between JJs $n$ and $n+1$. 
We make the ansatz 
\begin{IEEEeqnarray}{lCl}
 I_n^{o} &=& I_s^{o} + I_{\text{fl},n} \\
 I_n^{i} &=& I_s^{i} - I_{\text{fl},n}
\end{IEEEeqnarray}
for the currents on the outer and inner rail, 
where the screening currents $I_s^{o,i}$ are assumed to be homogeneous
and to have the same direction on both the outer and the inner rail.

From the quantization condition of the inner rail, 
 $L_i \sum_n I_n^{i} = 0$,
follows
\begin{equation}\label{eq:Is_inner_orig}
 I_s^i = \frac{1}{2 (N_u-1)} \sum_n I_{\text{fl},n} 
\end{equation}
where $N_u$ is the number of JJs in the upper (or lower) arm of the ring,
i.e. the total inductance of the inner LJJ rail amounts to 
$2(N_u-1) L_i$.
The outer rail is interrupted by two JJs, 
one each in the branch and the merge cell, with phases $\phi_{B,M}$ 
as indicated in Fig.~\ref{fig:boosterschematics}(b,c).
Once a fluxon (with polarity $\sigma = \pm 1$) has entered through the branch cell, 
$\phi_B \approx 2\pi \sigma$ while $\phi_M \approx 0$. 
The quantization condition on the outer rail therefore reads
$L_o \sum_n I_n^{o} = \Phi_0 (\phi_B - \phi_M)/(2\pi) \approx \sigma \Phi_0$, 
leading to 
\begin{equation}\label{eq:Is_outer_orig}
 I_s^o = \frac{\sigma \Phi_0}{2(N_u-1) L_o}  - \frac{1}{2 (N_u-1)} \sum_n I_{\text{fl},n} 
 \,.
\end{equation}
In \Eqs{eq:Is_inner_orig}, \eqref{eq:Is_outer_orig}
we see that the screening currents are directed in the same direction as 
the current contributed by the fluxon {\em on the outer rail}.

The current distribution of a Sine-Gordon fluxon centered at position $n_0$
is $I_{\text{fl},n} = 2\sigma I_c \lambda_J^2/(aW) \sech(a(n-n_0)/W)$, 
where $W=\lambda_J \sqrt{1-(v/c)^2}$ is the fluxon width.
(Here we adopted the convention that a fluxon with polarity $\sigma=1$
contributes CW-directed current on the outer rail, cf.~Fig.~\ref{fig:boosteroperations}.)
For sufficiently small discreteness we may estimate this 
in the continuum limit, $a/\lambda_J \to 0$, as
\begin{equation}
 \sum_n I_{\text{fl},n} \approx
 \frac{2\sigma I_c \lambda_J^2}{a^2} \int  \diff\theta \sech\theta 
 = \frac{\sigma \Phi_0}{L}
\end{equation}
where $L = L_o + L_i$.
The homogeneous screening currents thus read
\begin{eqnarray}
 I_s^i &=& \frac{\sigma \Phi_0}{2 (N_u-1) L} \\
 \label{eq:Is_outer}
 I_s^o &=&  \frac{\sigma \Phi_0}{2 (N_u-1) L} \frac{L_i}{L_o} 
 \,.
\end{eqnarray}
If the LJJ arms of the A-C ring are very large, such that $(N_u-1) L I_c/\Phi_0 = (N_u-1) \sqrt{a/\lambda_J}/(2\pi) \gg 1$, the screening currents are negligible. 
In small or moderately long LJJs, however, the screening currents can affect the dynamics near the branch and merge cell, where they implicitly bias
the termination JJs of the input and output LJJ. 
In the following, we qualitatively discuss several consequences 
of the screening currents on the fluxon dynamics in the ring LJJs connected 
to input and output LJJs.

\subsubsection{Entry at branch cell}
Because the generation of the screening currents costs energy, 
the quantization constraint creates a small potential barrier 
for a fluxon entering the ring from the input LJJ. 
And another effect contributes to this (small) potential barrier: 
the screening current on the outer rail 
creates an inherent current bias on the branch JJ (at $x_B$), 
which in turn exerts a force on the entering fluxon. 
This force is always repulsive because 
the screening currents generated by a fluxon (antifluxon) 
are directed CW (CCW). 
Therefore, in comparison with a very large booster ring (or the S-branch alone), 
the $\Ib$-threshold, which needs to be overcome for a 
fluxon to enter the ring, is shifted upwards. 

\subsubsection{Fluxon exit through merge cell}
If a fluxon is present in the A-C ring, 
the resulting screening current in the outer rail of the LJJ ring 
acts as an intrinsic bias on the output LJJ  
where it is connected to the LJJ ring at the merge point 
(and similarly on the input LJJ at the branch point). 
This bias affects the fluxon motion in such a way, 
that the fluxon scatters from the upper arm of the LJJ ring into the lower arm (where 
it is accelerated back towards the branch point).
To overcome this merging problem, 
special features in our booster designs are required.

\section{Alternative booster designs \texttt{D1J} -- \texttt{D3J}}
\label{app:Csh3_boosters}

Here we briefly summarize results for the booster designs of 
Fig.~\ref{fig:boosterschematics}(d).
In many ways, these boosters behave similar to those 
of Fig.~\ref{fig:boosterschematics}(a) such that many characteristics discussed
in Sec.~\ref{sec:dynamics} and Sec.~\ref{sec:performance}
apply to them too.

\begin{figure}[tb]
\includegraphics[width=\columnwidth]{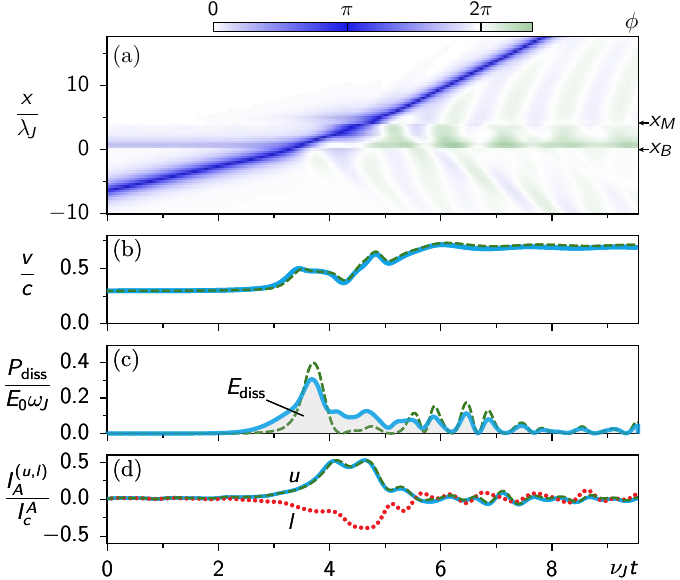}
\caption{
Simulated operation of booster $\DthrJ$ for $\Ib = 2.9 I_c$,
and an input fluxon with initial speed $v_i = 0.3c$. 
Panels (a-c) are similar to those of Fig.~\ref{fig:dynamics_SBL_Rs}, 
and panel (d) shows currents $I_A^{(u,l)}$
through upper and lower JJ bridges.
In (b-d) we compare the dynamics of $\DthrJ$, 
where two JJs have shunt resistors, with 
$R_{J,\text{crit}}/R_J^b = 0.4$ at $n_b$
and $R_{J,\text{crit}}/R_J^e = 0.2$ at $n_b+6$
(solid blue line), 
with that of $\DoneJ$, 
where only the JJ at $n=n_b$ is shunted, here with $R_{J,\text{crit}}/R_J^b = 0.6$ (dashed green).
} 
\label{fig:dynamics_SBL_Csh3_Rs}
\end{figure}

Figure~\ref{fig:dynamics_SBL_Csh3_Rs}
shows the dynamics for the booster design \texttt{D3J} of 
Fig.~\ref{fig:boosterschematics}(d), at $\Ib=2.9 I_c$.
Many aspects of the booster dynamics are similar to those of 
\texttt{D3}, cf.~Fig.~\ref{fig:dynamics_SBL_Rs}.
However, \texttt{D3J} does not have such large JJs in the merge cell 
as \texttt{D3} and thus avoids the pronounced slowing down of the fluxon 
near the merge cell. 
Instead, the fluxon velocity (panel b) increases almost monotonically with 
comparably minor retardations. 
This is made possible by the JJ bridges which temporarily store a part 
of the boost energy by which the fluxon is later powered upon reaching the merge cell. 
Panel (d) illustrates this process, showing  
the currents $I_A^{(u,l)}$ in the upper JJ bridge (solid) and lower JJ bridge (dashed).
In this design, the JJ bridges connect between the 6th JJ in each LJJ arm of the ring
and the 3rd JJ of the output LJJ. This allows a relatively good synchronization 
between the oscillation time of the JJ bridge 
and the time the fluxon needs to arrive at the merge cell.
The endpoint of the bridge connection to the output LJJ may be shifted closer to the merge 
cell, but we observed increased fluctuations in the A-C ring and bridge JJs in this case.

\begin{figure}[tb]
\includegraphics[width=\columnwidth]{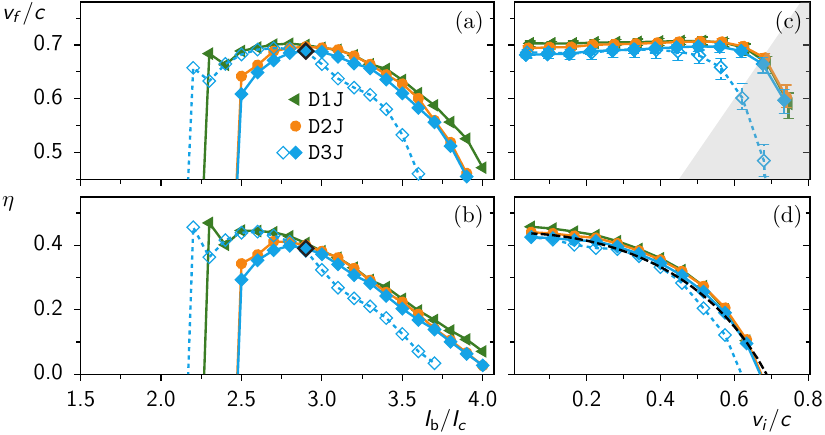}
\caption{
Booster performance, illustrated as in Fig.~\ref{fig:boosterperformance},
here for the three booster designs $\DoneJ, \DtwoJ, \DthrJ$ 
of Fig.~\ref{fig:boosterschematics}(d).
As in Fig.~\ref{fig:boosterperformance}, these use  shunt resistors 
$R_{J,\text{crit}}/R_J^b = 0.4$ and $R_{J,\text{crit}}/R_J^e = 0.2$ (filled markers) 
or 
$R_{J,\text{crit}}/R_J^b = 0.13$ and $R_{J,\text{crit}}/R_J^e = 0.07$ (open markers), respectively. 
In (a,b) the input velocity of the fluxon is set to $v_i = 0.3c$, 
for which a maximum in $v_f$ appears at around $\Ib=2.9 I_c$ 
in the final design $\DthrJ$ (emphasized marker). 
In (c,d) the bias is fixed at that value, $\Ib=2.9 I_c$.
}
\label{fig:boosterperformance_Csh3}
\end{figure}

Figure \ref{fig:boosterperformance_Csh3}
shows the performances of the booster designs $\DoneJ, \DtwoJ, \DthrJ$. 
These behave qualitatively similar to designs $\Done, \Dtwo, \Dthr$, 
cf.~Fig.~\ref{fig:boosterperformance_Csh3}, 
but have higher $\Ib$-thresholds and $v_f$ decreases faster at large $\Ib$-values. 
The maximum values of $v_f$ and $\eta$ are only slightly larger
compared with those of Fig.~\ref{fig:boosterperformance_Csh3}, 
cf.~also Table \ref{tab:boosterparameters}.
We can therefore conclude that the two design types have very similar 
performances, however, designs $\DoneJ, \DtwoJ, \DthrJ$ 
allow for a more monotonic fluxon acceleration with less retardation.

\section{Fluxon boost in biased LJJ}\label{app:boost_regularLJJ}

For comparison we analyze here the boosting action of a local bias current 
in a biased LJJ without branches or boundary effects. 
We assume that a bias current $\Ib$ is applied to a single JJ at position $x_b$ in the discrete LJJ.
The local damping coefficient 
$\alpha_b = R_{J,\text{crit}}/R_{J}^b = 1/(R_{J}^b C_J \omega_J)$
of that JJ may differ from that of other JJs, 
$\alpha = R_{J,\text{crit}}/R_J = 1/(R_J C_J \omega_J)$.
For $a/\lambda_J \ll 1$ (up to moderate values $\lesssim 1$), 
the dynamics may be approximated in the continuum limit by the perturbed SGE,
$\omega_J^{-2} \ddot \phi - \lambda_J^2 \phi'' + \sin\phi = f$,
here with the perturbation term
\begin{align}
f = -\frac{\alpha}{\omega_J} \dot \phi 
   + \left( \frac{\Ib a}{I_c}
   - \left(\alpha_b - \alpha \right) \frac{a}{\omega_J} \dot \phi 
   \right)
   \delta(x - x_b) 
   \,.
\end{align}
In the continuum limit, $(I_c, C_J, R_J^{-1}, L) \propto a$ go to zero,
whereas we assume that $R_J^b$ and $\Ib$ remain fixed when taking the limit 
$a/\lambda_J \to 0$, 
such that $\alpha_b a/\lambda_J$ is independent of $a$.
In contrast, 
the contribution of $\alpha a/\lambda_J$ vanishes
in the continuum limit, and the perturbation reduces to
\begin{align}
f = -\frac{\alpha}{\omega_J} \dot \phi 
   + \left( \frac{\Ib a}{I_c} - \frac{ \alpha_b a}{\omega_J} \dot \phi 
   \right)
   \delta(x - x_b) 
   \,.
\end{align}
Applying sine-Gordon perturbation theory \cite{McLaughlinScott1978},
under the assumption that parameters $\alpha$, $\Ib a/(I_c \lambda_J)$,
and $\alpha_b a/\lambda_J$ are sufficiently small, 
we obtain equations of motion for the position $X$ and velocity $v$
of a fluxon with polarity $\sigma$,
\begin{IEEEeqnarray}{lCl}
\IEEEyesnumber
\label{eq:sgpa}
\label{eq:sgpa_dX}
 \frac{1}{c} \dot X &=& \frac{v}{c} + f_X^{(\Ib)} + f_X^{(\alpha_b)} \IEEEyessubnumber\\ 
 \label{eq:sgpa_du}
 \frac{1}{\omega_J c}\dot v &=& -\alpha \frac{v}{c} \sqrt{1-\left(\frac{v}{c}\right)^2} + f_v^{(\Ib)} + f_v^{(\alpha_b)}  
 \IEEEyessubnumber
\end{IEEEeqnarray}
with dimensionless forces
\begin{align}
\label{eq:sgpa_forces_Ib}
\!\!\!f_X^{(\Ib)} &= \frac{f_v^{(\Ib)} \theta_b v/c}{\sqrt{1-(v/c)^2}} 
= \frac{\sigma \Ib a}{4 I_c \lambda_J} \frac{v}{c} \sqrt{1-\frac{v^2}{c^2}} \;\theta_b \sech\theta_b \\
\label{eq:sgpa_forces_alphab}
\!\!\!f_X^{(\alpha_b)} &= \frac{f_v^{(\alpha_b)} \theta_b v/c}{\sqrt{1-(v/c)^2}} 
= -\frac{\alpha_b a}{2\lambda_J} \frac{v^2}{c^2} \theta_b \sech^2\theta_b \\
\label{eq:thetab}
& \theta_b = \lambda_J^{-1} (x_b - X)/\sqrt{1- (v/c)^2}
\,.
\end{align}

\begin{figure}
\includegraphics[width=\columnwidth]{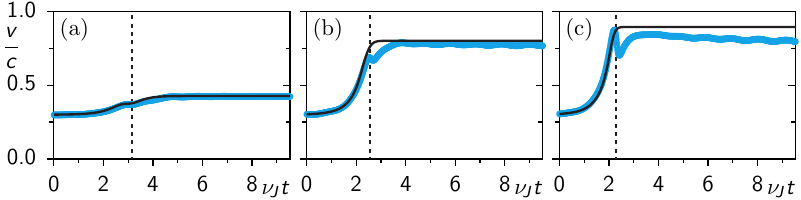}
\caption{
Fluxon velocity $v$ vs time $t$, for a fluxon in a regular (discrete) LJJ 
with a single current-biased JJ, 
for (a) $\Ib/I_c = 0.5$, (b) $\Ib/I_c = 2.8$ and (c) $\Ib/I_c = 5.0$.
We shunt the biased JJ with a resistor 
$R_{J}^b = 1/(\alpha_b C_J \omega_J) = R_{J,\text{crit}}/\alpha_b$, 
with $\alpha_b a/\lambda_J = 0.6/\sqrt{7}$, 
whereas all other JJs are undamped, $\alpha=0$.
From the simulated LJJ dynamics, 
with LJJ discreteness of $a/\lambda_J = 1/\sqrt{7}$ (blue markers),
one obtains $v=\dot X$ after fitting the LJJ phases $\phi_n(t)$
to a fluxon phase distribution centered at position $X$. 
The simulation is compared with the perturbation dynamics (black line),
\Eq{eq:sgpa}. 
For (a) small bias  there is reasonable agreement, 
but at (b,c) larger bias the perturbation dynamics fails to predict the characteristic `dip' of the fluxon velocity just after passing the biased JJ
(indicated by dashed vertical line).
}
\label{fig:LJJ_fluxonboost}
\end{figure}

Figure \ref{fig:LJJ_fluxonboost} illustrates the boost dynamics 
for three different $\Ib$-values,
comparing the dynamics obtained from \Eq{eq:sgpa} (black lines), 
with that from the circuit simulation (blue markers).
For small bias (panel a) there is reasonable agreement between the two, 
but at larger bias (panels b,c) the perturbation dynamics fails to predict the characteristic `dip' of the fluxon velocity just after passing the biased JJ
at $x_b$ (time of passage indicated by dashed vertical line).
We ascribe this phenomenon to the presence of a finite-width localized mode which is sustained by the bias current and whose amplitude scales with $\Ib$, 
see the paragraph below. 
In interaction with this mode, the fluxon typically undergoes two distinct 
boost phases, respectively caused by the attractive (repulsive) force 
before (after) the mode center. 
This interpretation is supported by a 
collective coordinate model of the fluxon dynamics where we include a localized mode and which then qualitatively reproduces (not shown) 
the features seen in the circuit simulation.  

Even at small bias (panel a), 
a small retardation in the fluxon acceleration is observable, 
in the fluxon dynamics
obtained from either the circuit simulation or the perturbation equations. 
This effect can be attributed to the different length scales of the forces, 
\Eqs{eq:sgpa_forces_Ib} and \eqref{eq:sgpa_forces_alphab}, 
where the damping forces $f_{X,v}^{(\alpha_b)}$ are stronger confined 
around $x_b$ than the boost forces $f_{X,v}^{(\Ib)}$,
and thus can temporarily slow down (or even invert) the acceleration,
depending on the values of $v$, $\Ib$, and $\alpha_b$.
Note that at large bias (panels b,c) the total damping force on the fluxon 
is larger in the circuit simulation compared with \Eq{eq:sgpa}.
This increased damping is a result of the finite discreteness 
and the related energy loss \cite{WusOsb2020_RFL, BraKiv1998}.


Before moving on to evaluate the potential energy step and dissipation energy
for the LJJ in SGP approximation, 
let us briefly sketch the estimates for the above-mentioned localized mode amplitude: 
Using the ansatz
$\phi_{\text{loc}} = A e^{\mu |x-x_b|/\lambda_J}$ 
and inserting it into the linearized Lagrangian of the continous LJJ
one obtains the equation of motion 
$\omega_J^2 \ddot A + (\mu^2 + 1) A = \mu \Ib a/(I_c \lambda_J)$
for the amplitude $A$, 
resulting in the steady-state amplitude
$A = \mu (\mu^2 + 1)^{-1} \Ib a/(I_c \lambda_J)$.
The dimensionless  width $\mu$ can be estimated from the dispersion relation 
of the bulk SGE for the steady state, 
$0 = \omega_J^2 + 2 c^2/a^2 \left( 1 - \cosh(a\mu/\lambda_J)\right)$, 
leading to $\mu = (\lambda_J/a) \text{Arcosh}\left(1 + a^2/(2\lambda_J^2)\right) \to 1$,
where the continuum limit $a/\lambda_J \to 0$ is taken. 

\subsection{Potential energy step generated by $\Ib$}
\label{app:Epot_Ib}

Here we set $\alpha=0$, $\alpha_b=0$ and $\Ib \neq 0$.
The localized conservative force $f_v^{(\Ib)}$ in \Eq{eq:sgpa} accelerates a fluxon (antifluxon) to the right (left). 
In order to calculate the height of the corresponding potential energy step,
we start from equation $\Efl(v)$ for the fluxon energy, \Eq{eq:Efl},
whose time derivative is 
\begin{equation}\label{eq:dEfl_dt}
\dot \Efl
= \frac{8 E_0 v \dot v}{c^2(1-(v/c)^2)^{3/2}}
= \frac{8 E_0 \omega_J v f_v}{c(1-(v/c)^2)^{3/2}} 
\;.
\end{equation}
When evaluating $\dot \Efl$ for the local force $f_v^{(\Ib)}$ 
of \Eq{eq:sgpa_forces_Ib}, 
we can make use of the property
\begin{eqnarray}\label{eq:dthetaB_dt_localperturbation}
 \dot \theta_b
   &=& \frac{\theta_b v \dot v/c^2}{1-(v/c)^2} - \frac{\dot X/\lambda_J}{\sqrt{1-(v/c)^2}} 
=-\frac{v/c}{\sqrt{1-(v/c)^2}}
\,
\end{eqnarray}
which follows from \Eq{eq:thetab}
after inserting \Eq{eq:sgpa} 
and using the relation
between $f_X^{(\Ib)}$ and $f_v^{(\Ib)}$
given in \Eq{eq:sgpa_forces_Ib}.
(Note that this special relation between $f_X$ and $f_v$
holds for any {\em local} SGE-perturbations.)
Thus, in the case of $f_u^{(\Ib)}$,  \Eq{eq:dEfl_dt}  evaluates to 
\begin{align}
\dot \Efl
 &=  2 E_0 \frac{\sigma \Ib a}{I_c \lambda_J} \frac{v/c}{\sqrt{1-(v/c)^2}} \sech\theta_b   \nonumber 
 = -2 E_0 \frac{\sigma \Ib a}{I_c \lambda_J} \dot \theta_b  \sech\theta_b \\
 &=  -2 E_0 \frac{\sigma \Ib a}{I_c \lambda_J}\frac{\diff}{\diff t} 
 \left( 2 \arctan e^{\theta_b} \right) \,.
\end{align}
Integrating this equation one obtains
the fluxon energy difference before and after the 
scattering, 
\begin{align}
\Delta \Efl 
= \int_{t_i}^{t_f} \diff t \dot \Efl
 = \frac{2\pi \sigma \Ib a}{I_c \lambda_J}  E_0
 = \sigma \Phi_0 \Ib
 \,,
\end{align}
and thus the height of the potential energy step $E_{\text{pot}} = -\Delta \Efl$ created by $\Ib$ is 
\begin{equation}\label{eq:Epot_Ib}
 E_{\text{pot}} = -\sigma \Ib \Phi_0
 \,.
\end{equation}

From the energy conservation 
$\Efl(t_i) = \Efl(t_f) +  E_{\text{pot}}$ we can now determine 
the asymptotic fluxon velocity $v_f = v(t_f)$ after the scattering, 
 \begin{equation}\label{eq:uf_Ib}
  \frac{v_f}{c} = \sqrt{1 - \frac{1-(v_i/c)^2}{ \left(1 + \frac{\pi \sigma}{4} \frac{\Ib a}{I_c \lambda_J} \sqrt{1-\frac{v_i^2}{c^2}} \right)^2 }}
 \end{equation}
where $v_i=v(t_i)$ is the initial fluxon velocity.


The particular form of $f_{X,v}^{(\Ib)}$ also allows to directly integrate \Eq{eq:sgpa}, such that we can determine not only the step height, 
but the potential as function of the fluxon position $X$ and velocity $v$.
To do so, we transform \Eq{eq:sgpa_du} into an equation of motion for 
the fluxon momentum $P = 8 E_0 v c^{-2} \left(1 - (v/c)^2\right)^{-1/2}$,
and integrating it we find the Hamiltonian 
\begin{equation}\label{eq:Hamiltonian_localbias}
 H_{\text{fl}}(X,P) = \Efl(v)
 - 4 E_0 \frac{\Ib}{I_0} \arctan e^{-\sigma \theta_b}
\end{equation}
with the kinetic energy contribution 
$\Efl(v) = \frac{8 E_0}{\sqrt{1-(v/c)^2}}=\sqrt{(8E_0)^2 + (P c)^2}$.
From \Eq{eq:Hamiltonian_localbias} we can determine the potential step height
as
\begin{eqnarray}
 E_{\text{pot}} = 
 \left[- 4 E_0 \frac{\Ib a}{I_c \lambda_J} \arctan e^{-\sigma \theta_b}\right]^{\theta_b=-\infty}_{\theta_b=\infty}
\end{eqnarray}
leading to the same result as \Eq{eq:Epot_Ib}.

\subsection{Dissipation generated by $\alpha_b$}
\label{app:Ediss_alphab}

Here we set $\alpha=0$, $\Ib=0$ and $\alpha_b=0$.
The localized force $f_v^{(\alpha_b)}$ in \Eq{eq:sgpa} acts as a damping. 
For a fluxon that passes the damped JJ, 
we approximate the energy loss due to that
damping force, following similar steps as in Sec.~\ref{app:Epot_Ib}. 
Again one can again make use of the property \Eq{eq:dthetaB_dt_localperturbation}
since $f^{(\alpha_b)}$ is a localized perturbation.
Starting with \Eq{eq:dEfl_dt}, with $f_v^{(\alpha_b)}$ from \Eq{eq:sgpa_forces_alphab},
 \begin{align}
\dot \Efl
 &= -\frac{4 E_0 \omega_J \alpha_b a}{\lambda_J} \frac{v^2/c^2}{(1-(v/c)^2)} \sech^2\theta_b \nonumber \\
 &= -\frac{4 E_0 \alpha_b a}{\omega_J \lambda_J} \dot \theta_b^2 \sech^2\theta_b 
 = -\frac{4 E_0 \alpha_b a}{\omega_J \lambda_J} \dot \theta_b
\frac{\diff \left(\tanh\theta_b\right)}{\diff t} \nonumber \\
\label{eq:dEfl_dt_localRJ_0}
 &= -\frac{4 E_0 \alpha_b a}{\omega_J \lambda_J} 
\left( \frac{\diff}{\diff t} \left[\dot \theta_b \tanh\theta_b \right]
- \ddot \theta_b \tanh\theta_b \right)
\end{align}
Taking the derivative of \Eq{eq:dthetaB_dt_localperturbation},
\begin{eqnarray*}
\ddot \theta_b = -\frac{\dot v}{\lambda_J (1-(v/c)^2)^{3/2}} 
  = -\frac{f_v^{(\alpha_b)}}{\lambda_J (1-(v/c)^2)^{3/2}}
  \,,
\end{eqnarray*}
we see that it contributes with a higher perturbation order in \Eq{eq:dEfl_dt_localRJ_0}, 
$\ddot \theta_b / \dot \theta_b \propto \alpha_b a/\lambda_J$,
and we therefore neglect it here. 
For the energy $\Ediss = - \Delta \Efl$ dissipated during the passage of the damped JJ this approximation then yields
\begin{align}
\label{eq:Ediss_alphab}
\!\!\!\Ediss
\approx 
\frac{4 E_0\alpha_b a}{\lambda_J}
  \left(\!  \frac{|v_f|/c}{\sqrt{1-(v_f/c)^2}} 
        + \frac{|v_i|/c}{\sqrt{1-(v_i/c)^2}} \!\right) .
\end{align}
From the energy balance $\Efl(t_f) =  \Efl(t_i)-\Ediss$
then follows the asymptotic fluxon velocity $v_f$ after the scattering:
\begin{equation}\label{eq:uf_RJb}
  \frac{v_f}{c} = \frac{-\lambda + \sqrt{\zeta_0(\zeta_0 - 1 +\lambda^2)}}{\lambda^2 + \zeta_0}
\end{equation}
with $\lambda = \alpha_b a/(2\lambda_J)$
and $\zeta_0 = (1 - \lambda v_i/c)^2 / (1-(v_i/c)^2)$.

\subsection{Fluxon boost for $\Ib \neq 0$  and $\alpha_b \neq 0$}

\begin{figure}[t]
\includegraphics[width=\columnwidth]{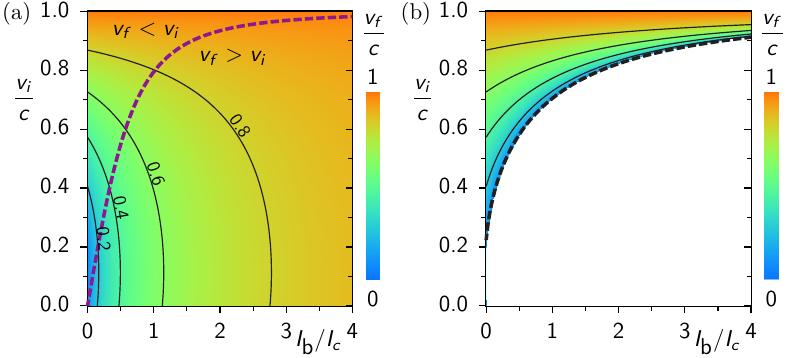}
\caption{
 Velocity $v_f$ after scattering of fluxon with initial speed $v_i$
 at local current bias $\Ib$, 
 for  (a) $\sigma=1$ and (b) $\sigma=-1$, 
 according to the analytic approximation, \Eq{eq:uf_Ib_RJb}. 
 The biased JJ is also resistively shunted 
 ($\alpha_b a/\lambda_b = 0.6/\sqrt{7}$),
 as can be noted in the lines of constant $v_f$ at $\Ib=0$, where $v_f < v_i$.
 (a) For $\sigma=1$ the fluxon velocity is boosted, i.e. $v_f>v_i$, 
 for $v_i \leq \text{max}(v_i)$, 
 with $\text{max}(v_i)$ from \Eq{eq:fluxonboost_criterion} (purple dashed line).
 (b) For $\sigma=-1$ the initial velocity $v_i$ needs to be 
 above the threshold $v_{i,\text{th}}$,
 \Eq{eq:ui_th_Ib_RJb} (black dashed line), 
 for the fluxon to be transmitted over the biased point. 
}
\label{fig:SGPAanalysis_vf_eta}
\end{figure}

Here we set $\alpha=0$, but $\alpha_b \neq 0$ and $\Ib \neq 0$.
With the potential energy step and the dissipation energy from 
\Eqs{eq:Epot_Ib} and \eqref{eq:Ediss_alphab}
one can analyze the energy balance 
$\Efl(t_f) - \Efl(t_i) = -E_{\text{pot}} -\Ediss$
in presence of both perturbations
leading to the asymptotic fluxon velocity
\begin{align}\label{eq:uf_Ib_RJb}
 & \frac{v_f}{c} = \frac{-\lambda + \sqrt{\zeta(\zeta - 1 +\lambda^2)}}{\lambda^2 + \zeta} \\
 & \text{with}\quad 
  \zeta = \frac{\left(1 - \lambda \frac{v_i}{c} + \frac{\pi \sigma \Ib a}{4 I_c \lambda_J}\sqrt{1-(v_i/c)^2}\right)^2}{1-(v_i/c)^2} \nonumber
  \,.
\end{align}
The asymptotic velocity $v_f$ is shown in Fig.~\ref{fig:SGPAanalysis_vf_eta}
as function of $\Ib$ and $v_i$ and $\sigma$. 

For the case $\sigma\Ib > 0$ (panel a), 
where the potential forms a downward step, 
a fluxon boost, $v_f > v_i$, is always possible as long as $\Ib$ is sufficiently large. 
The upper-velocity limit, below which the fluxon is boosted 
can be calculated by setting $v_f = v_i$ in the energy balance,
resulting in the boost criterion
\begin{equation}\label{eq:fluxonboost_criterion}
v_i \leq \text{max}(v_{i}) 
 = c \left( 1 + \left(\frac{4 \alpha_b a/\lambda_J}{\pi\sigma \Ib a/(I_c\lambda_J)} \right)^2 \right)^{-1/2}
\end{equation}
This value is indicated in panel (a)
by the purple line, below which boosting occurs.

The efficiency of the fluxon boost is
defined as the ratio of the fluxon energy gain $\Efl(v_f) - \Efl(v_i)$
to the energy cost. 
The energy cost is the work done by the current source to 
generate the potential of height $E_{\text{pot}}$ for the fluxon, 
and thus the boost efficiency is
\begin{equation}\label{eq:eta_SGPA}
\!\!\eta := \frac{  \Efl(v_f) - \Efl(v_i)  }{  -E_{\text{pot}}  }  
       = 1 + \frac{\Ediss}{ E_{\text{pot}} }
       = 1 - \frac{\Ediss}{ \sigma \Ib \Phi_0}
\end{equation}
Herein, the dissipation energy is given by \Eq{eq:Ediss_alphab},
which can be evaluated with $v_f$ from \Eq{eq:uf_Ib_RJb} 
as a function of $\Ib, \alpha_b$ and $v_i$ alone. 

For the case $\sigma\Ib < 0$ (panel b), where the potential forms an upward step,
only a sufficiently fast fluxon can pass the biased JJ. 
Setting $v_f = 0$ in the energy conservation,
the corresponding threshold velocity is found,
 \begin{equation}\label{eq:ui_th_Ib_RJb}
 \frac{v_{i,\text{th}}}{c} = \frac{\lambda + \sqrt{\chi(\chi - 1 +\lambda^2)}}{\lambda^2 + \chi}
\end{equation}
with $\chi = \left(1 - \pi \sigma \Ib a/(4 I_c \lambda_J) \right)^2$.
This threshold value is indicated in panel (b) by the black line.
 
\end{appendix}


\end{document}